\documentclass[twocolumn]{aastex61}
\pdfoutput=1 
\usepackage{amsmath,amstext}
\usepackage[T1]{fontenc}
\usepackage{apjfonts} 
\usepackage[figure,figure*]{hypcap}
\usepackage{gensymb}
\usepackage{natbib}

\shorttitle{W49A Draft}
\shortauthors{De Pree et al.}

\begin{document}

\title{Time-Variable Radio Recombination Line Emission in W49A}


 
\author[0000-0003-3115-9359]{C. G. De Pree}
\affil{Department of Physics \& Astronomy, Agnes Scott College, Decatur, GA, United States}
\author[0000-0003-1526-7587]{D. J. Wilner}
\affil{Center for Astrophysics | Harvard \& Smithsonian, Cambridge, MA, United States}
\author[0000-0003-1159-3721]{L. E. Kristensen}
\affil{Niels Bohr Institute \& Centre for Star and Planet Formation,  Copenhagen University, Denmark}
\affil{Center for Astrophysics | Harvard \& Smithsonian, Cambridge, MA, United States}
\author[0000-0003-1480-4643]{R. Galv\'an-Madrid}
\affil{Instituto de Radioastronom\'{\i}a y Astrof\'{\i}sica (IRyA), UNAM, Morelia, Michoac\'an, Mexico}
\author{W. M. Goss}
\affil{National Radio Astronomy Observatory, Socorro, NM, United States,}
\author[0000-0002-0560-3172]{R. S. Klessen}
\affil{Universit\"{a}t Heidelberg, Zentrum f\"{u}r Astronomie,
  Institut f\"{u}r Theoretische Astrophysik,
Heidelberg, Germany}
\affil{Universit\"{a}t Heidelberg, Interdisziplin\"{a}res Zentrum f\"{u}r Wissenschaftliches Rechnen, Heidelberg, Germany}
\author[0000-0003- 0064-4060]{M.-M. Mac Low}
\affil{Department of Astrophysics, American Museum of Natural History, New York, NY, USA}
\affil{Center for Computational Astrophysics, Flatiron Institute, New York, NY, United States}
\author{T. Peters}
\affil{Max-Planck-Institut f\"{u}r Astrophysik, Garching, Germany}
\author{A. Robinson}
\affil{Department of Physics \& Astronomy, Agnes Scott College, Decatur, GA, United States}
\author{S. Sloman}
\affil{Department of Physics \& Astronomy, Agnes Scott College, Decatur, GA, United States}
\author{M. Rao}
\affil{Department of Physics \& Astronomy, Agnes Scott College, Decatur, GA, United States}

\begin{abstract}
We present new Jansky Very Large Array (VLA) images of the central region of the W49A star-forming region at 3.6~cm and at 7~mm
at resolutions of 0\farcs15 (1650 au) and 0\farcs04 (440 au), respectively. 
The 3.6~cm data reveal new morphological detail in the ultracompact \ion{H}{2} region population, as well as several previously 
unknown and unresolved sources. In particular, source
A shows elongated, edge-brightened, bipolar lobes, indicative of a collimated outflow, and source
E is resolved into three spherical components. 
We also present VLA observations of radio recombination lines at 3.6~cm and 7~mm, and IRAM Northern Extended Millimeter Array (NOEMA) observations at 1.2~mm. 
Three of the smallest ultracompact \ion{H}{2} regions (sources
A, B2 and G2) all show broad kinematic linewidths,  with $\Delta$V$_{FWHM}\gtrsim$40~km~s$^{-1}$.  A multi-line analysis indicates  that broad linewidths remain after correcting for pressure broadening effects, 
       suggesting
the presence of supersonic flows. 
Substantial changes in linewidth over the 21 year time baseline at both 3.6 cm and 7 mm are found for source G2. 
At 3.6 cm, the linewidth of G2 changed from 
31.7$\pm$1.8 km s$^{-1}$ to 55.6$\pm$2.7 km s$^{-1}$, an increase of $+$23.9$\pm$3.4 km s$^{-1}$.
The G2 source was
    previously
reported to have shown a 3.6~cm continuum flux density decrease of 40\% between 1994 and 2015.  
This source sits near the center of a very young bipolar outflow whose variability may have produced these changes. 
\end{abstract}

\section{INTRODUCTION}


  Regions of ionized hydrogen (\ion{H}{2} regions) are detected in the vicinity of young, high mass stars. Some of the earliest high resolution radio observations of star forming regions found very small \ion{H}{2} regions associated with the earliest stages of star formation, with diameters an order of magnitude smaller than the canonical $\sim$1 pc \citep{wood89}. These ultracompact (UC) \ion{H}{2} regions have typical diameters of $\sim$0.1 pc and electron densities of n$_e\sim$10$^4$ cm$^{-3}$, while hypercompact (HC) \ion{H}{2} regions are even smaller (d$\sim$0.03 pc) and higher density (n$_e\sim$10$^6$ cm$^{-3}$) regions (Kurtz et al. 2005). Additionally, \cite{yang2018} point out that HC \ion{H}{2} regions also differ from UC \ion{H}{2} regions in that they typically have broader radio recombination line widths, ranging from 40-100 km s$^{-1}$ instead of the typical 25-30 km s$^{-1}$ found in most UC \ion{H}{2} regions. \cite{r-s2020} discuss the definition of  HC \ion{H}{2} regions, and the possibility that the term `hypercompact' should be used to refer to regions based on their size alone. In this paper, we use the existing definition of  HC \ion{H}{2} regions from the literature, and as a result, the regions discussed in this paper are referred to as ultracompact (UC) \ion{H}{2} regions. 
  
  Recognition of the prevalence and significance of short timescale variability in UC and HC \ion{H}{2} regions has been growing on both the
observational and theoretical fronts.
   It has been understood for decades that hot molecular cores are rotating
   and infalling \citep{keto1990}, and that their central massive protostars drive
   bipolar outflows, one less than 400 years old \citep{mac1994}.

Detailed numerical models suggest that some of the known unusual characteristics and morphologies of UC and 
HC \ion{H}{2} regions could be caused by the trapping of material in accretion flows that form massive stars \citep{peters2010a},
   and that the bipolar outflows seen emerging from massive star-forming regions
   could be formed by the combination of mutiple jets from these regions
   \citep{peters2014}.
    These models
predict that compact ionized regions around very young
massive stars can vary significantly in radio flux density over years to decades
\citep{gm2011}. Observationally, a number of UC and HC \ion{H}{2} regions have been 
discovered to show flux density variations on these timescales
\citep{acord1998, her2004, vtd2005, gm2008, dzib2013, dep2014, dep2015, riv2015, hunter2018, bro2018, dep2018}.

Our own work has focused on the Sgr B2 and W49A star forming regions, each of which harbor a large number of UC~\ion{H}{2} and HC~\ion{H}{2} regions highly 
clustered within a small area on the sky that provide an ensemble that can be relatively easily monitored for variability.
We have reported flux density changes over a $\sim$20 year time span for several sources within the 
Sgr B2 Main and North regions \citep{dep2014, dep2015}, and more recently in the source W49A/G2 \citep{dep2018}. 
We note, however, that most of the radio sources in these highly clustered regions exhibited no significant changes,
    also demonstrating the time stability of the flux calibrations. 

Notably, in W49A the
G2 source that decreased at 3.6~cm by 20\% in peak intensity (from 71$\pm$4 to 57$\pm$3 mJy
beam$^{-1}$), and 40\% in integrated flux (from 109$\pm$11 to 67$\pm$7 mJy), is located
   within
the highest velocity water maser outflow 
in  the Galaxy \citep{gmr1992,mcgrath2004}. A water maser flare was detected in this region in 2014 \citep{kra2015}, just prior to the latest epoch of radio observations.

The 3D hydrodynamic models of \cite{peters2010a, peters2010b} indicate that many UC and HC \ion{H}{2} regions
     should be
associated with bipolar outflows 
of ionized gas.
Targeted observations of ionized gas at the highest frequencies and angular resolutions available--made with the Karl G. Jansky Very Large Array\footnote{The National Radio Astronomy Observatory is a facility of the National Science Foundation operated under cooperative agreement by Associated Universities, Inc.} (VLA)  and the NOrthern Extended Millimeter Array (NOEMA)--may be especially instructive.
The free-free continuum emission at higher frequencies has lower optical depth, and pressure broadening of radio recombination lines 
becomes negligible, presenting an opportunity to reveal the kinematics in the innermost regions of these sources.

In this paper, we present new VLA observations of the W49A region
(distance $11.1^{+0.8}_{-0.7}$~kpc, \cite{zhang2013}). In \cite{dep2018}, we presented only the B-configuration continuum 3.6 cm data from the new observations, in order to look for flux density changes since the mid-1990s.  At 3.6 cm, the VLA data presented here include A-configuration continuum, and combined B-, C-, and D-configuration 
continuum and line observations. This mode of combining the data allowed for direct comparison to VLA data taken in 1994-5, when no A-configuration data were taken. We also present VLA A-configuration continuum and line observations at 7 mm. In addition, NOEMA 7A6~configuration observations were made at 1.2~mm. 
Section 2 describes the VLA and NOEMA observations. 
Section 3 presents the new radio continuum observations and associated recombination line observations. 
Section 4 discusses possible connections between the line detections and the previously reported flux density variations,
and the need for regular, high resolution, temporal monitoring of the compact ionized regions to better assess the variable accretion
hypothesis. Section 5 summarizes the conclusions. 

\section{OBSERVATIONS AND DATA REDUCTION}

\subsection{VLA 3.6 cm}
We observed W49A with the VLA at 3.6 cm through the cycle of 
the B, A, D and C configurations in 2015--2016.  
The setup consisted of bands centered at 8.5 and 9.716 GHz, with 8 contiguous sub-bands at each frequency. 
To make a direct comparison of these data with 3.6~cm observations from the 1994--1995 epoch \citep{dep1997}, we focus in this paper on the 8.5 GHz band. 
The H91$\alpha$, H92$\alpha$ and H93$\alpha$ lines were each covered in narrow sub-bands with 16 MHz bandwidth ($\sim$560 km s$^{-1}$) with 128 
channels, each covering 125 kHz (4.5 km~s$^{-1}$). The other 5 wide sub-bands had a 128 MHz bandwidth with 128 (1~MHz) channels. The H92$\alpha$ line is at 8.309382 GHz. Phase and flux density calibrations were carried out with the Common Astronomy Software Applications (CASA) pipeline \citep{mcmullin2007}. Calibrator sources used are listed in Table 1. Standard pipeline calibrations (optimized for continuum observations) at first flagged the recombination lines as interference, so the CASA pipeline was run a second time with the known recombination line frequencies marked so that the channels containing line emission were not flagged.

\subsubsection{Continuum}
\begin{figure*}[t!]
\gridline{\fig{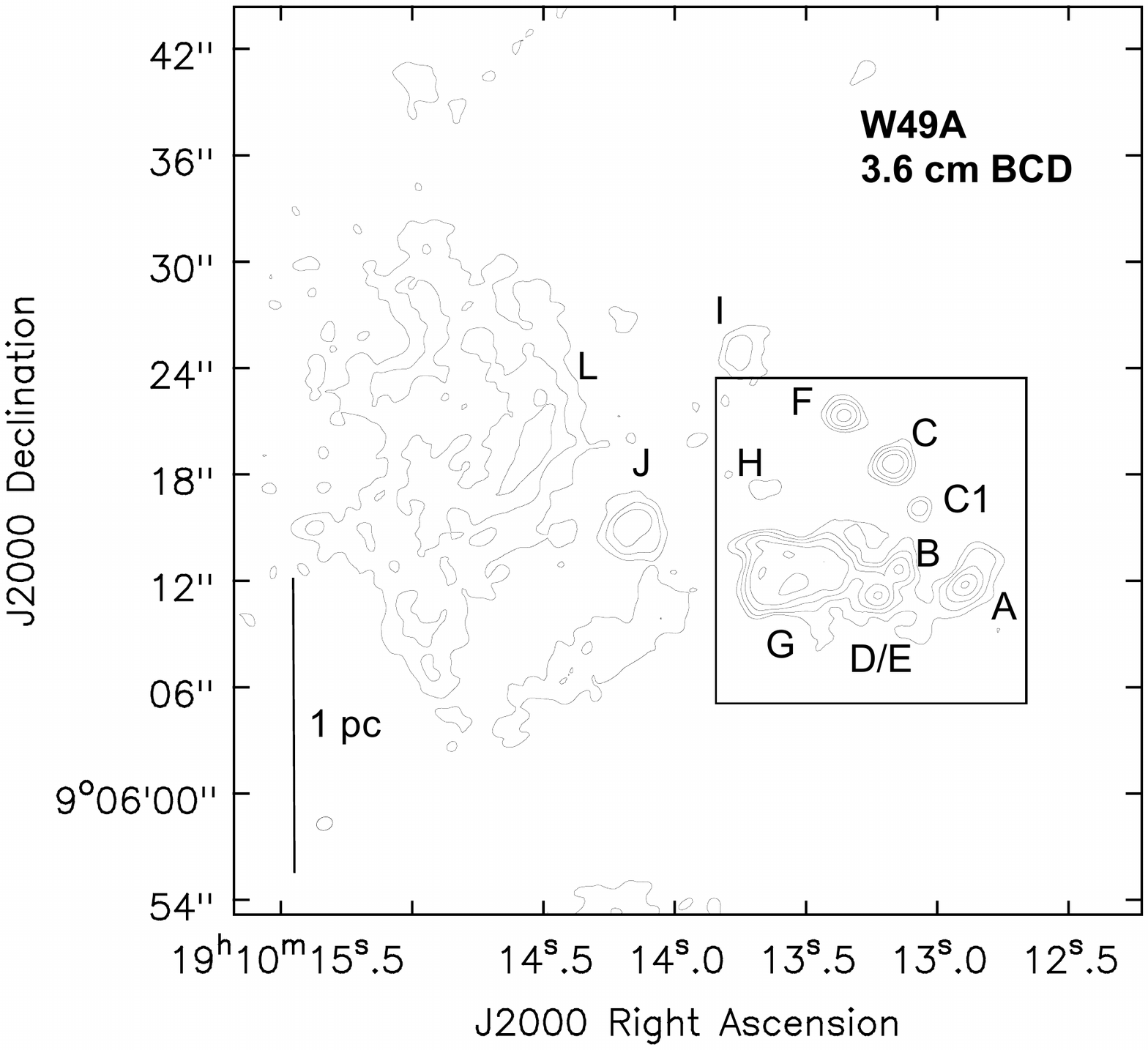}{0.75\textwidth}{}
        }
\vspace{-5.0cm}
\caption{VLA 3.6~cm continuum image of W49A from the combined BCD configurations, with beam size $\theta_{beam}\sim$0\farcs8.
Contours are -1, 1, 2, 4, 8 and 16 $\times$ 5$\sigma$ (4.0~mJy beam$^{-1}$). 
Sources are labeled as in \cite{dep1997} and \cite{dep2000}. The vertical bar indicates 1 pc at 11.1 kpc. 
The box indicates the region highlighted in Figures~\ref{fig:3.6cmA} and \ref{fig:7mmA}. }
\label{fig:3.6cmBCD} 
\end{figure*}

\begin{figure*}[t!]
\gridline{\fig{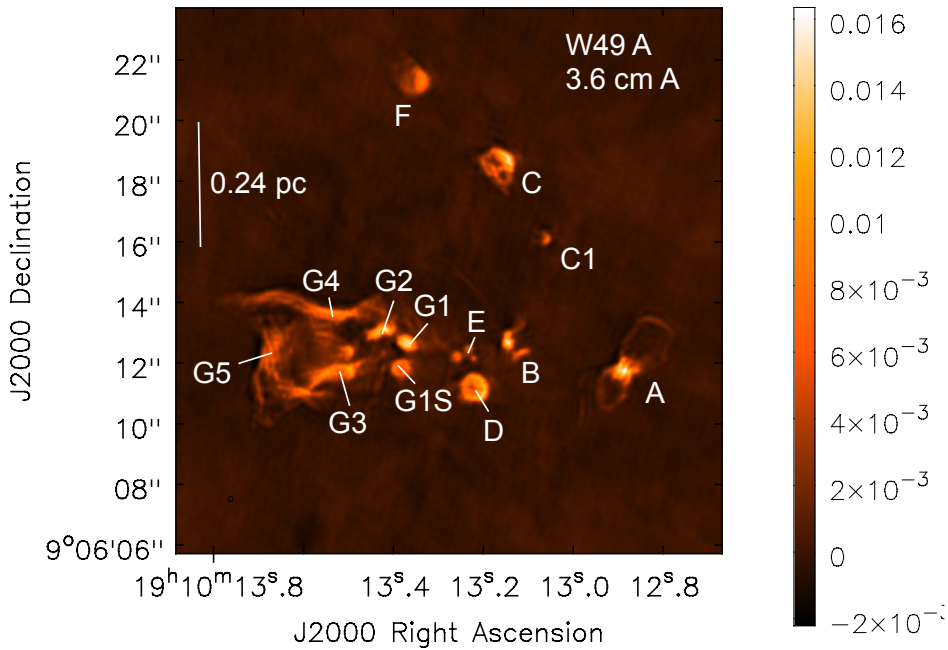}{0.75\textwidth}{}
        }
\vspace{-7.0cm}
\caption{VLA 3.6 cm continuum image of W49A from the A-configuration, with beam size $\theta_{beam}\sim$0\farcs15.
The wedge indicates source brightness in mJy beam$^{-1}$, and the vertical bar indicates 0.24 pc.}
\label{fig:3.6cmA} 
\end{figure*}

For the observations from the B, C, and D configurations, the 5 wide sub-bands at 8.5 GHz were combined together in the CASA package, imaged and self-calibrated (3 phase only cycles, with solution intervals of 90s, 60s and 30s) to produce a single continuum image with 0\farcs92$\times$0\farcs74 resolution.
Figure~\ref{fig:3.6cmBCD} shows the continuum image from the combined B, C, and D configurations (hereafter BCD).
The observations from the A configuration were processed in the same way, with 3 self-calibration cycles using the same solution intervals.
Figure~\ref{fig:3.6cmA} shows this higher resolution image, with synthesized beam 0\farcs16$\times$0\farcs15, for the 
central region of W49A (corresponding approximately to the $\sim$1\arcmin~FWHM primary beam of the 7~mm image). 
Table~\ref{tab:VLAobs} lists additional specifics of the resulting continuum images.

\begin{deluxetable*}{lccccc}
\tablecaption{Observational Parameters}
\tablewidth{700pt}
\tabletypesize{\scriptsize}
\tablehead{\colhead{Parameter} & \colhead{7 mm (VLA-A)} & \colhead{3.6 cm (VLA-A)} & \colhead{3.6 cm (VLA-BCD)} & \colhead{NOEMA} \\
(1) & (2) & (3) & (4) &(5)}
\startdata
Date (Configuration) & Sep 27 2016 (VLA-A) & Jun 24 2015 (VLA-A) & Feb 08 2015 (VLA-B) & Aug 20, 21, 24 2015 (NOEMA-7A6) \\
 & & & Jan 29 2016 (VLA-C) & \\
  & & & Oct 10 2015 (VLA-D) & \\
Observing Program & 16B-022 & 15A-089 & 15A-089 & S15-AR\\
Total Observing Time (hr) & 4 & 3  & 3
   (Each) & 9.75 \\
Number of Antennas & 27 & 27 & 27 (All) & 6 \\
Beam FWHM (arcsec, BPA) & & \\
\it{Line } & 0.060 $\times$ 0.040, -79$\degree$ & N/A\tablenotemark{a}  & 0.95$\times$0.75, -66$\degree$ & 0\farcs$58\times$0\farcs22, 18\degree \\
\it{Continuum} & 0.044 $\times$ 0.036, -73$\degree$ &  0.16 $\times$ 0.15, -4$\degree$  & 0.92$\times$0.74, -69$\degree$ & 0\farcs$56\times$0\farcs23, 19\degree\\
RMS Noise (mJy beam$^{-1}$) & & \\  
\it{Line } &  0.2  & NA\tablenotemark{a}  & 0.5  & 50 \\
\it{Continuum} & 0.16  &  0.13  & 0.8  & 24 \\
Radio LSR central velocity (km s$^{-1}$) & 8.0 & 8.0 & 8.0 & 7.0\\
Summed Bandwidth(MHz) & 896\tablenotemark{b} & 688\tablenotemark{c}  & 688\tablenotemark{c} & 4000 \\
Number of channels (per IF) & 128 &  NA & 128 & 1024 \\
Channel separation (kHz, km s$^{-1}$) & 500 (3.3) & NA &  125 (4.5) & 3900 (4.6)\\
Flux density calibrator & 3C 286 & 3C 286 & 3C 286 (All) & MWC349, 1749+096\\
Phase calibrator & J1922+1530 & J1925+2106 & J1925+2106 (All) & 1749+096 \\
Bandpass calibrator & J1922+1530 & J0319+4130 & J0319+4130 (All) & 1749+096\\
Time on source (hr) & 1.5 & 1.75 & 1.75 (B), 1.9(C), 1.75 (D)  & 7.1\\
Rest frequency (GHz) & 45.454 & 8.3094 & 8.3094 & 257 GHz\\
\enddata
\tablenotetext{a}{Recombination lines were not detected in A configuration data alone.}
\tablenotetext{b}{Bandwidth values are for the 7 bands without a recombination line.}
\tablenotetext{c}{Bandwidth values are for the 5 bands without a recombination line}
\label{tab:VLAobs}
\end{deluxetable*}

\subsubsection{Radio Recombination Lines}
In total, 6 recombination lines were covered in these observations: H86$\alpha$, H87$\alpha$, H88$\alpha$ (at 9.716 GHZ), and H91$\alpha$, H92$\alpha$, and H93$\alpha$ (at 8.5 GHz). In this paper, we analyze only the H92$\alpha$ line emission, since that was the only line observed in the 1994-5 VLA observations. The sub-band containing the H92$\alpha$ line data was imaged and self-calibrated for each configuration (A, B, C, D), using the solution intervals noted above. 
At the high angular resolution of the A configuration alone, brightness sensitivity limitations precluded detection of H92$\alpha$ line 
emission from any of the individual sources.
Indeed, in their theoretical investigation of recombination line observations with the VLA, \cite{peters2012} found that A-configuration-only observations would not have sufficient sensitivity, even in full-synthesis observations.

We therefore proceeded to make a lower resolution image of the H92$\alpha$ line by combining the narrow band containing the H92$\alpha$ line from the 
B, C, and D configurations. These data were imaged and self-calibrated as described above, using the same solution intervals. The continuum was subtracted to produce a line-only image of the H92$\alpha$ emission. This line image has synthesized beam resolution of 0\farcs95$\times$0\farcs75.  Like the 3.6~cm continuum
imaging, the line imaging at this resolution closely matches the previously published H92$\alpha$ data \citep{dep1997}, facilitating a direct comparison. The other recombination lines were not observed in the previous epoch, and will be published separately.

\begin{figure*}[t!]
\gridline{\fig{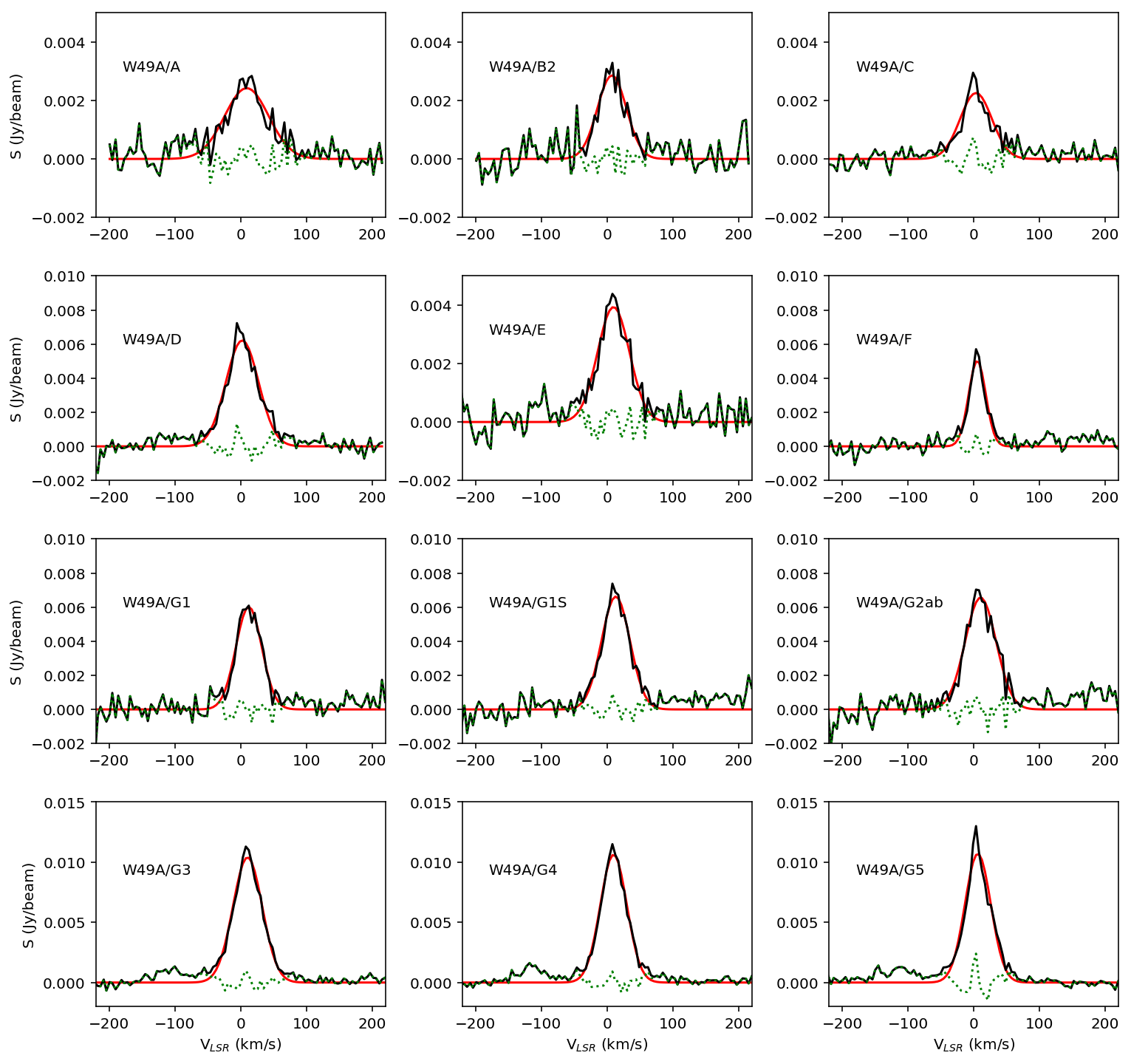}{0.75\textwidth}{}
        }
\caption{Sources with detected H92$\alpha$ line emission in the central region of W49A. 
The data are shown as solid black lines, Gaussian fits as solid red lines, and residuals in dotted green lines. 
The rms noise in the BCD H92$\alpha$ line data is 0.5 mJy beam$^{-1}$. 
The source name (indicated in Figure~\ref{fig:3.6cmA}) is indicated on each line profile.}
\label{fig:H92alpha} 
\end{figure*}

Gaussian  fits were made to each of the detected 3.6 cm sources using the SciPy Python library. Figure~\ref{fig:H92alpha} shows the line profiles from the BCD imaging of the sources with H92$\alpha$ detections, together with Gaussian fits. Table~\ref{tab:RRL} lists the corresponding fit parameters to these recombination lines. 

These new data have $\sim$30\% higher sensitivity than those of the earlier epoch of \citet{dep1997}, 
and the VLA correlator upgrade provides substantially more bandwidth, nearly doubling it from $\sim$220 km s$^{-1}$ to $\sim$420 km s$^{-1}$ in the current observations. In addition, the previous observations were centered between the H92$\alpha$ and He92$\alpha$ lines, placing the H92$\alpha$ close 
to the band edge.
We note, for example, that the broad H92$\alpha$ line not previously detected toward source A is readily apparent in 
these new observations.

\begin{deluxetable*}{lcccccccc}[t!]
\tablecaption{Radio Recombination Line Parameters of Central Sources of W49A}
\tablewidth{700pt}
\tabletypesize{\scriptsize}
\tablehead{
\colhead{Source Name} & \colhead{Recombination} & \colhead{Amplitude} & \colhead{V$_{LSR}$}  & \colhead{$\Delta V_{\rm Current}$} & \colhead{$\Delta V_{\rm Archival}$}\tablenotemark{a} & Ratio  & \colhead{$\Delta V_{\rm Kin}$} & \colhead{n$_{e}$}  \\ 
\colhead{} & \colhead{Line} & \colhead{(mJy beam$^{-1}$)} &\colhead{(km s$^{-1}$)} & \colhead{(km s$^{-1}$)} & \colhead{(km s$^{-1}$)} & (Current/Arch) &\colhead{(km s$^{-1}$)} &  \colhead{(10$^5$ cm$^{-3}$)}\\
(1) & (2) & (3) & (4) &(5) & (6) & (7) & (8) & (9)
}
\startdata
\hline\hline
A 
& H52$\alpha$ & 1.2$\pm$0.1  & 13.2$\pm$0.7& 43.6$\pm$1.7 & 46.8$\pm$2.2 & 0.9$\pm$0.1  & 39.2 & 1.7 \\
  & H92$\alpha$ & 2.4$\pm$0.1 & 9.3$\pm$2.1& 78.9$\pm$5.1 &  ND &  N/C \\
  \hline
B$_{1}$ &H52$\alpha$ & 2.1$\pm$0.2  & 6.1$\pm$2.4 & 48.3$\pm$5.6 & N/D & N/C & 44.4\\
 & H92$\alpha$ & N/D & N/D & N/D & N/C & N/C \\
B$_{2}$
&H52$\alpha$ & 1.2$\pm$0.1  & 15.9$\pm$0.8 & 49.5$\pm$2.0 & 48.7$\pm$2.8 & 1.0$\pm$0.1 & 45.6 & 0.2 \\
 & H92$\alpha$ & 2.9$\pm$0.2 & 7.2$\pm$1.8 & 51.9$\pm$4.3 & 56.4$\pm$5.4 & 0.9$\pm$0.1  \\
 \hline
C & H52$\alpha$ & 0.74$\pm$0.04  & -9.5$\pm$0.8 & 31.42$\pm$2.0 & 39.3$\pm$2.4 & 0.8$\pm$0.1 & 24.9 & 1.7\\
 & H92$\alpha$ & 2.3$\pm$0.1 &3.4$\pm$1.4 & 57.0$\pm$3.3 & 18.6$\pm$2.8 & 3.1$\pm$0.5\tablenotemark{b} &  \\
 \hline
D & H52$\alpha$ & 0.95$\pm$0.3 & 4.7$\pm$0.5 & 33.1$\pm$1.1 & 35.1$\pm$1.2 & 0.9$\pm$0.1 & 27.0 & 1.7 \\
 & H92$\alpha$ & 6.2$\pm$0.2 &2.5$\pm$0.8 & 58.7$\pm$1.9 & 48.1$\pm$1.6 & 1.2$\pm$0.1  \\
 \hline
E1 & H52$\alpha$ & 1.0$\pm$0.1 & 7.4$\pm$2.7 & 43.6$\pm$6.36 & N/D & N/C & N/C & N/C \\
E & H92$\alpha$ & 3.9$\pm$0.2 & 9.7$\pm$1.3 & 57.3$\pm$3.1 & N/D & N/C   \\
\hline
F & H52$\alpha$  & N/D & N/D & N/D & 26.2$\pm$1.4 & N/C & N/C \\
 & H92$\alpha$ & 5.0$\pm$0.2 & 5.3$\pm$0.7 & 30.8$\pm$1.6 & 28.1$\pm$1.2 &  1.1$\pm$0.1\\
\hline
G1 & H52$\alpha$ & 2.1$\pm$0.1  & 2.4$\pm$0.4 & 28.6$\pm$1.0 & 29.2$\pm$2.9\tablenotemark{c} & 1.0$\pm$0.1 & 21.3 & 1.2 \\
  & H92$\alpha$ & 6.0$\pm$0.2 & 12.5$\pm$1.0 & 46.5$\pm$2.4 & 29.3$\pm$2.0 & 1.6$\pm$0.1 \\
G1 South & H52$\alpha$ & 0.7$\pm$0.1 & 10.8$\pm$1.0 & 33.6$\pm$2.5 & N/D & N/C & 27.6 & 1.2 \\
  & H92$\alpha$ & 6.6$\pm$0.3 & 13.1$\pm$1.0 & 51.2$\pm$2.3 & N/D & N/C \\
\hline
G2a & H52$\alpha$ & 2.1$\pm$0.1  & 5.6$\pm$1.7 & 62.4$\pm$4.0 & 40.1$\pm$1.1\tablenotemark{d} & 1.6$\pm$0.1 & 59.4 & N/C \\
G2b & H52$\alpha$ & 2.4$\pm$0.1  & 19.9$\pm$1.3& 54.7$\pm$3.2 & 40.1$\pm$1.1\tablenotemark{d} & 1.4$\pm$0.1 & 51.3& N/C \\
G2c & H52$\alpha$ & 1.1$\pm$0.1  & 12.9$\pm$4.0 & 69.6$\pm$9.4 & 40.1$\pm$1.1\tablenotemark{d} & 1.7$\pm$0.2 & 66.9 & N/C \\
G2ab & H92$\alpha$ & 6.6$\pm$0.3 & 10.4$\pm$1.2 & 55.6$\pm$2.7 & 31.7$\pm$1.8\tablenotemark{e} & 1.8$\pm$0.1 \\
\hline
G3a & H52$\alpha$ & 0.9$\pm$0.01  & 16.6$\pm$2.7 & 34.3$\pm$6.3 &34.3$\pm$0.4\tablenotemark{f} & 1.0$\pm$0.2 & 21.6 & 1.3 \\
G3b & H52$\alpha$ & 1.8$\pm$0.01  & 14.1$\pm$1.1 & 27.2$\pm$2.5 &34.3$\pm$0.4\tablenotemark{f} & 0.8$\pm$0.1 & \\
G3c & H52$\alpha$ & 1.4$\pm$0.1  & 10.0$\pm$0.6 & 31.0$\pm$1.3 &34.3$\pm$0.4\tablenotemark{f} & 0.9$\pm$0.1 & \\
G3d & H52$\alpha$ & 1.5$\pm$0.1  & 3.2$\pm$0.7 & 25.3$\pm$1.7 &34.3$\pm$0.4\tablenotemark{f} & 0.7$\pm$0.1 & \\
G3  & H92$\alpha$ & 10.4$\pm$0.2 & 10.7$\pm$0.5 & 52.7$\pm$1.2 & 35.9$\pm$1.0 & 1.5$\pm$1.0 \\
\hline
G4 & H52$\alpha$ & 1.7$\pm$0.1  & 7.8$\pm$0.4 &23.5$\pm$1.0 & 34.3$\pm$0.4\tablenotemark{f} & 0.7$\pm$0.1 & 13.7 & 1.5 \\
  & H92$\alpha$ & 10.6$\pm$0.2 & 9.8$\pm$0.5 & 47.6$\pm$1.2 & 42.3$\pm$0.9 & 1.1$\pm$0.1 \\
\hline
G5 & H52$\alpha$ & N/D & N/D & N/D & N/D & N/C & N/C & N/C \\
  & H92$\alpha$ & 10.7$\pm$0.3 & 6.2$\pm$0.6 & 46.5$\pm$1.4 & 38.1$\pm$1.1 & 1.2$\pm$0.1 \\
\hline
\enddata
\tablenotetext{a}{Archival linewidth measurements for H92$\alpha$ and H52$\alpha$ from \cite{dep1997}}
\tablenotetext{b}{H92$\alpha$ line is a borderline detection in the archival data with a S/N at $\sim$4$\sigma$ for W49A/C in \cite{dep1997}. Archival linewidth value for W49A/C is likely an underestimate. We note that the ratio of the H52$\alpha$ lines is consistent with no change.}
\tablenotetext{c}{Archival G1 line parameter taken from \cite{dep2004}}
\tablenotetext{d}{Here we use the H52$\alpha$ linewidth reported for G$_{West}$ from \cite{dep1997}}
\tablenotetext{e}{Here we use the H92$\alpha$ linewidth reported for G2 from \cite{dep1997}}
\tablenotetext{f}{Here we use the H52$\alpha$ linewidth reported for G$_{East}$ from \cite{dep1997}}
\label{tab:RRL}
\end{deluxetable*}

\subsection{VLA 7 mm}
We observed W49A with the VLA at 7 mm on September 27 2016 in the A configuration. 
The setup consisted of bands centered at 45.453 and 48.153 GHz, with 8 contiguous sub-bands at each frequency. 
The H52$\alpha$ line (45.453719 GHz) was covered in a sub-band of width 64 MHz ($\sim$420 km s$^{-1}$) with 
128 channels of width 500 kHz (3.3 km s$^{-1}$). 
The other 7 sub-bands had 128 MHz bandwidth with 64 (2~MHz) channels. 
Due to interference and calibration issues with the 48.153 GHz band, only the 45.454 data were fully reduced. Additionally, the new H52$\alpha$ observations can be directly compared to the same recombination line observed in 1995 \citep{dep1997}.
As with the 3.6 cm data, phase and flux density calibrations were carried out with the CASA pipeline. Calibrator sources used are listed in Table 1. Similar flagging issues caused us to run the pipeline a second time with the known recombination line frequencies marked.

\subsubsection{Continuum}
\begin{figure*}[t!]
\gridline{\fig{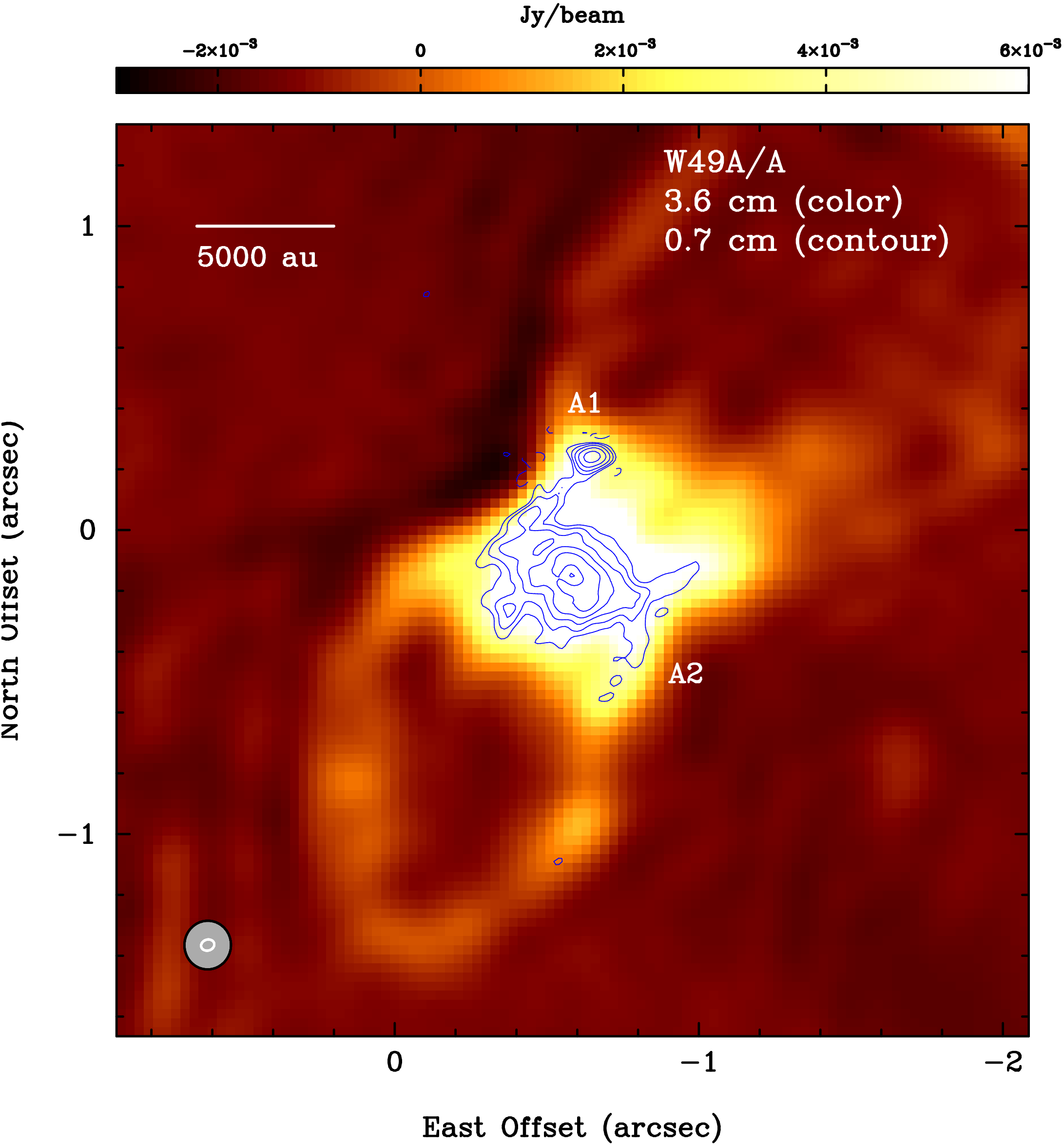}{0.3\textwidth}{(a)}
          \fig{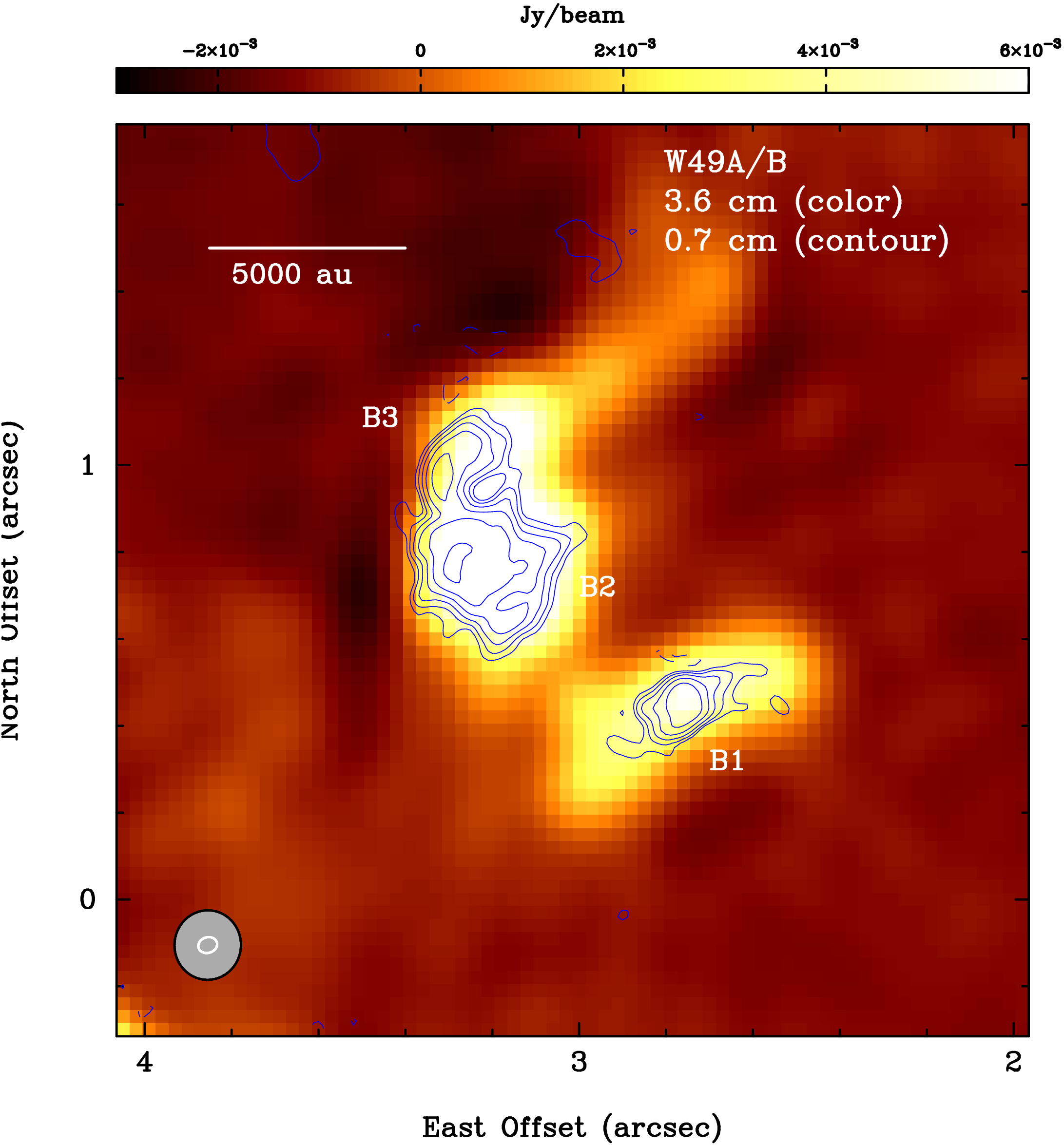}{0.3\textwidth}{(b)}
          \fig{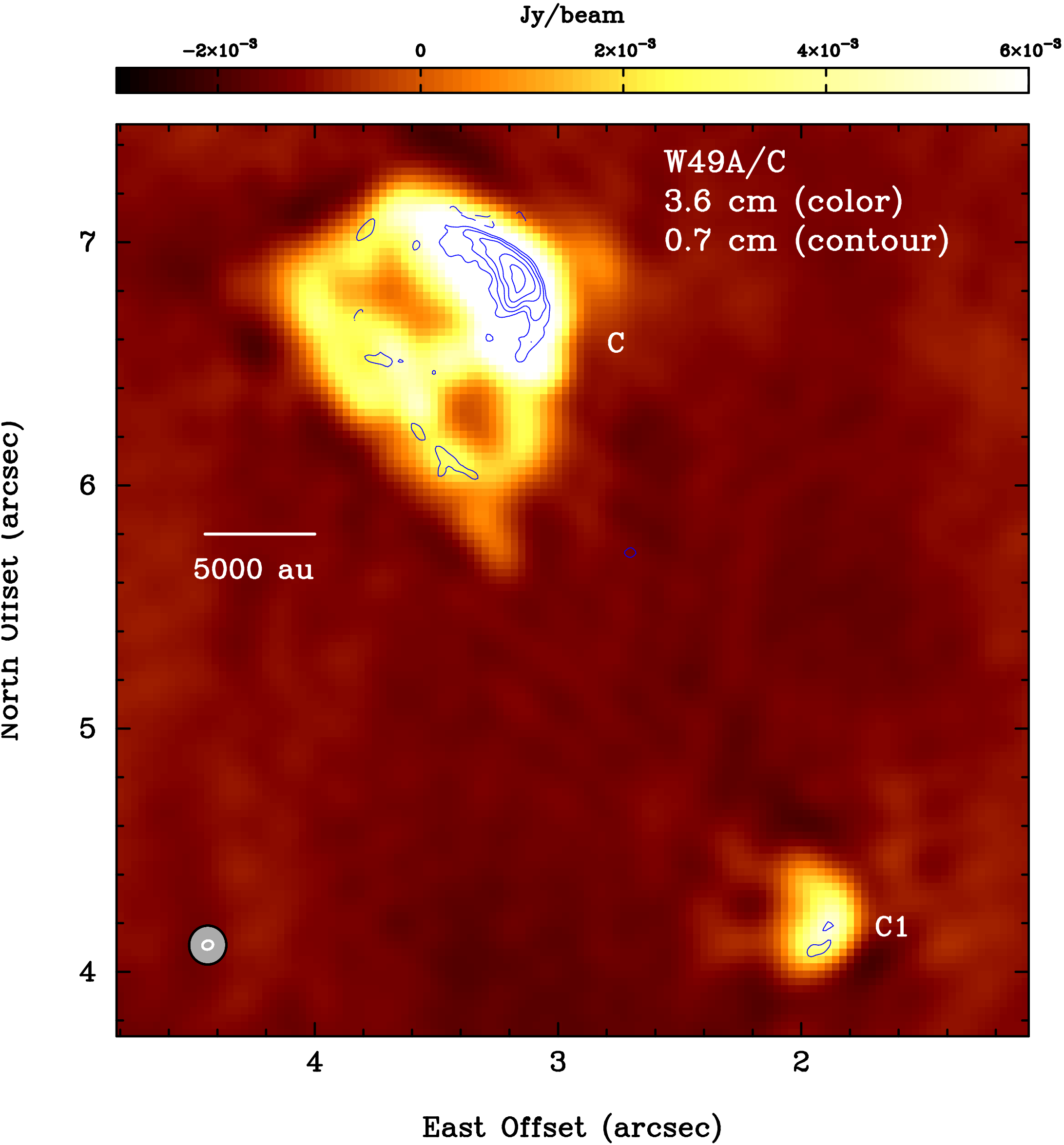}{0.3\textwidth}{(c)}
          }
\gridline{\fig{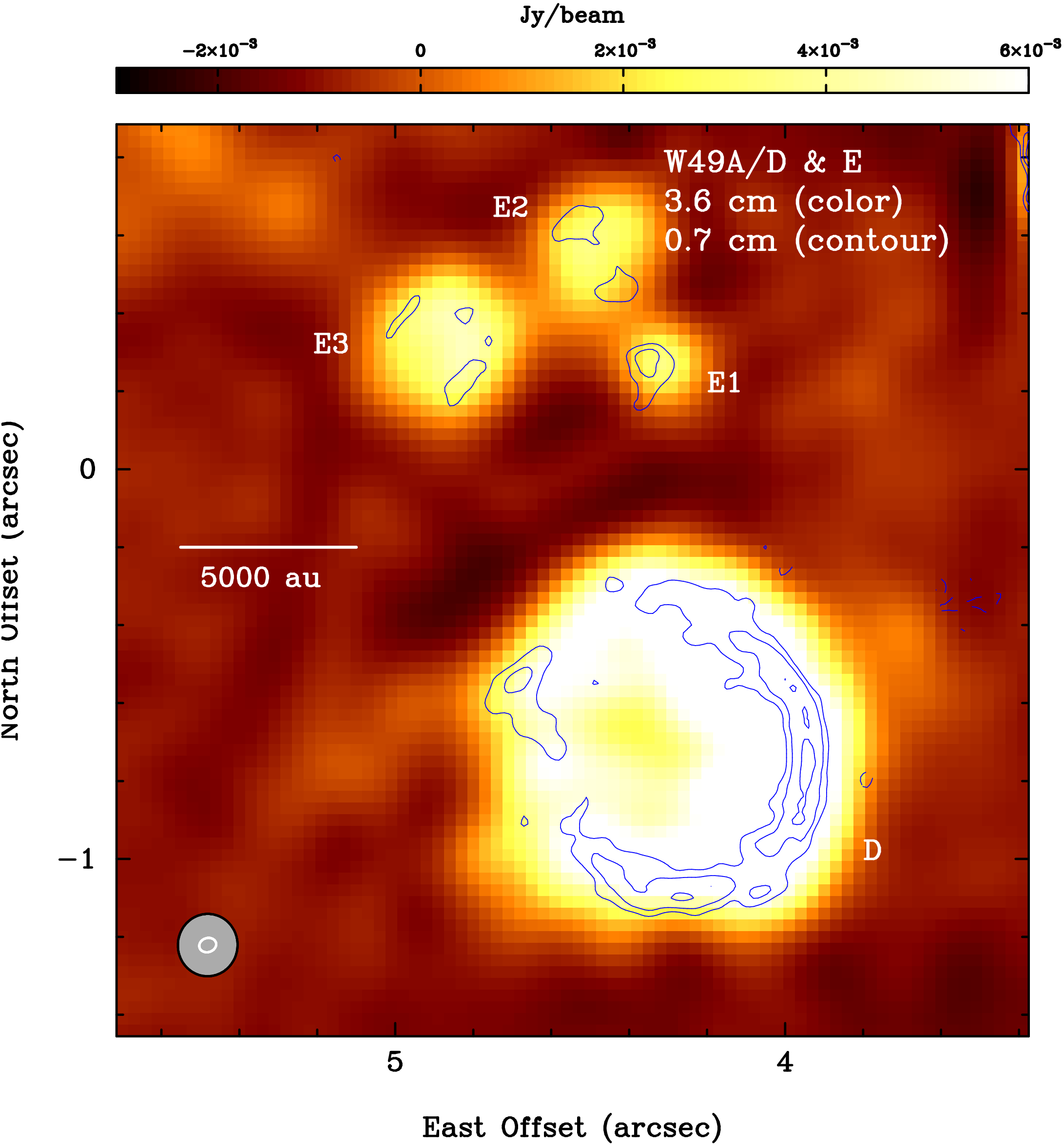}{0.3\textwidth}{(d)}
          \fig{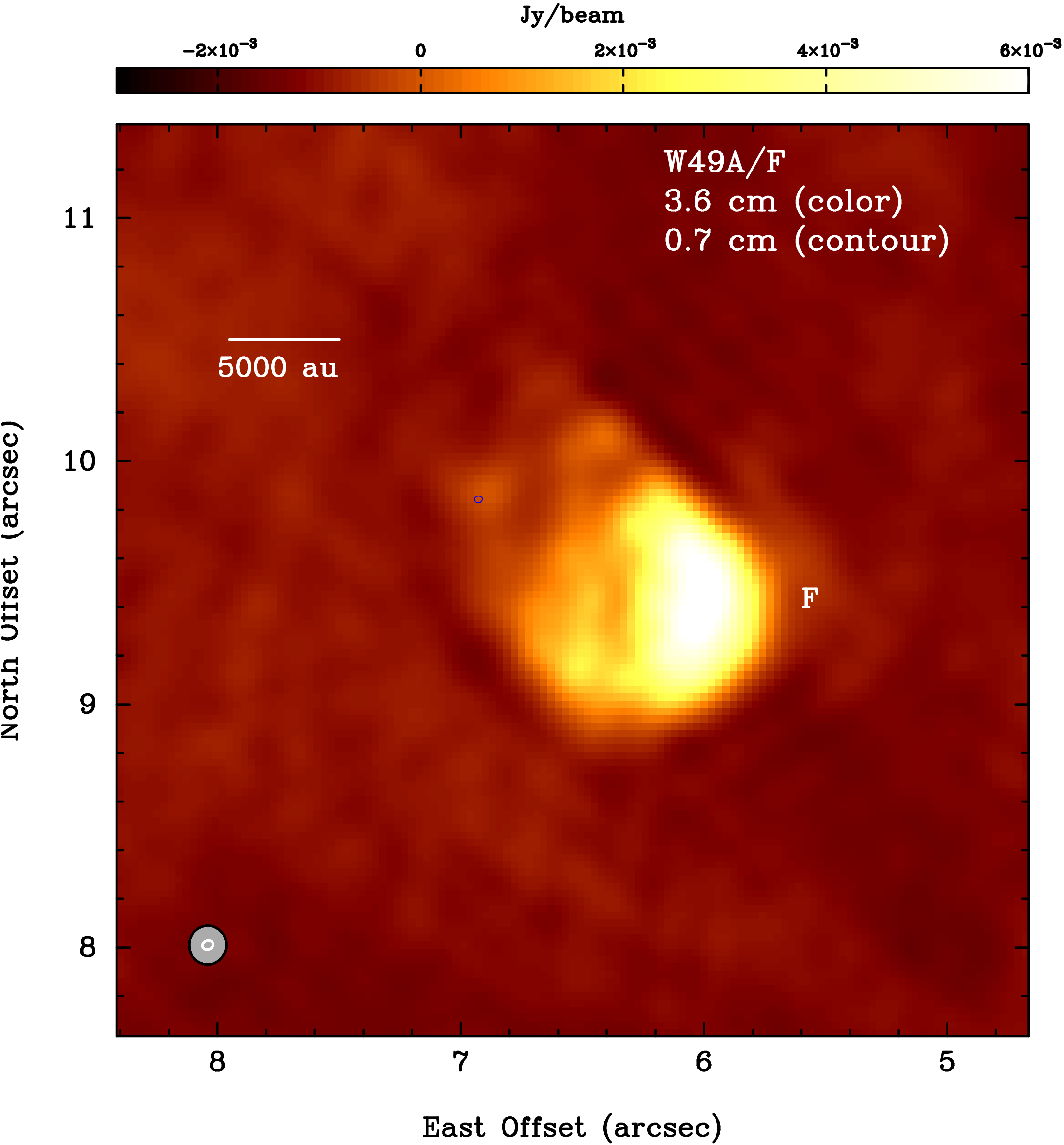}{0.3\textwidth}{(e)}
          \fig{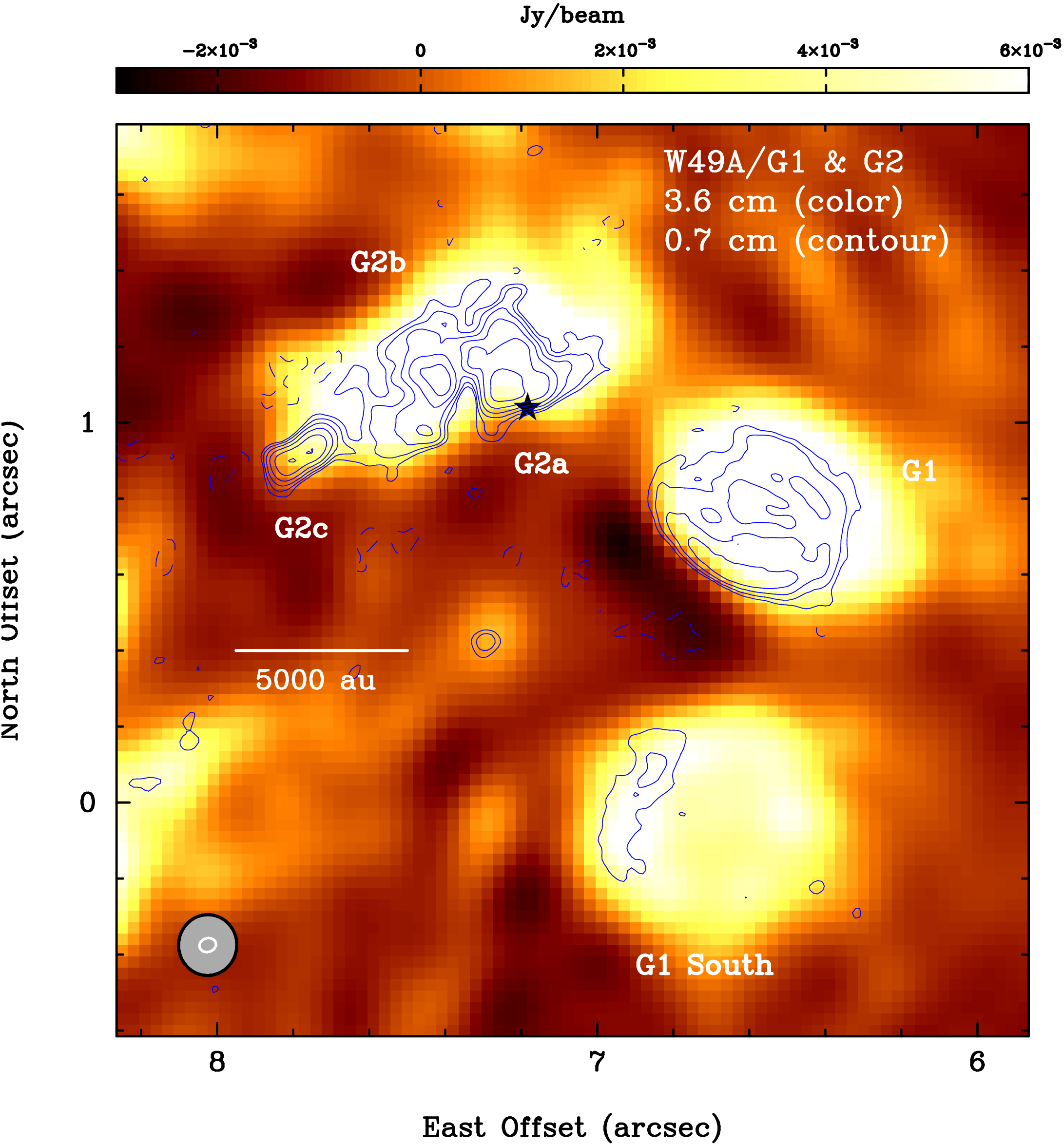}{0.3\textwidth}{(f)}
          }
\gridline{\fig{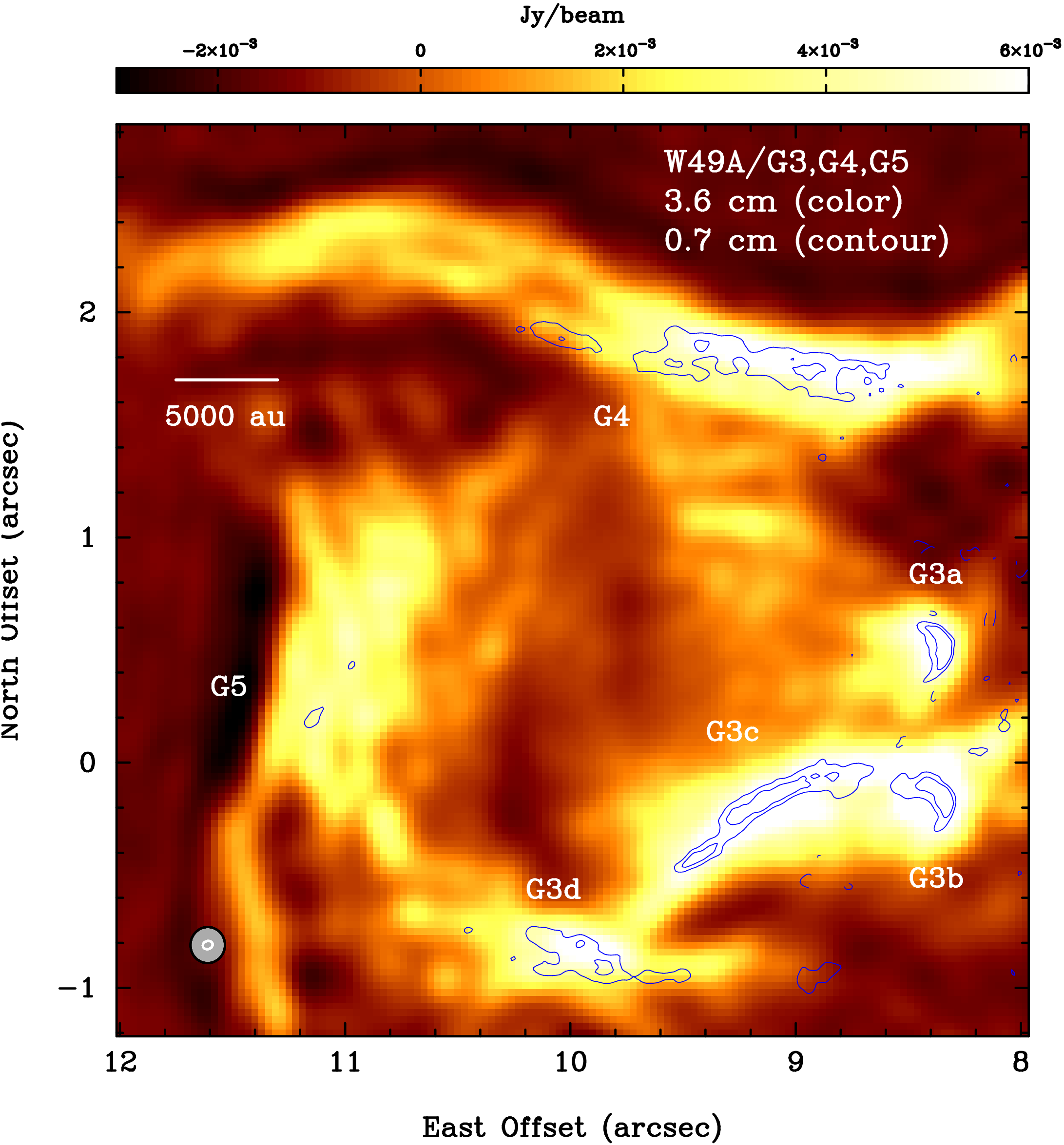}{0.3\textwidth}{(g)}
          }
\caption{Detailed images from the 3.6 cm and 7 mm A configuration data. The 3.6 cm images ($\theta_{beam}\sim$0\farcs15) are shown in pseudocolor and the 7 mm images ($\theta_{beam}\sim$0\farcs04) are shown as blue contours. 
The contours levels are -1, 1, 2, 4, 8, 16 and 32 $\times 0.8$~mJy beam$^{-1}$ (5$\sigma$).
Sources in (a) - (g) are labeled as in \cite{dep1997} and \cite{dep2000}. 
The horizontal bar corresponds to 5000 au in each image. 
The synthesized beams for the 3.6 cm and 7 mm images are indicated by the gray and white ellipses in the lower left corner of each panel. Offsets in arcsec are from the pointing center of both observations: RA 19h10m12.93s, Dec +09$\degree$06$\arcmin$11.882$\arcsec$. The black star in (f) indicates the center of expansion of the water masers, as reported in \cite{gmr1992}.}
\label{fig:7mmA} 
\end{figure*}

The continuum data from the 7 wide sub-bands were imaged and self-calibrated using three cycles of phase-only self-calibration, using solution intervals of 150s, 120s and 90s. 
Figure~\ref{fig:7mmA} shows the final 7~mm continuum image (as blue contours) overlaid on the corresponding 3.6~cm A-configuration sources, with synthesized beam size of 0\farcs044$\times$0\farcs036. 

At the resolution of the A-configuration image at both frequencies, the sources identified in Figure~\ref{fig:3.6cmA} are resolved into the named subsources.
Table~\ref{tab:VLAobs} provides additional information about these 7 mm observations and image properties.

To provide measures of the flux densities and sizes of the detected 7~mm sources, we fitted them with 2D Gaussians, where applicable, 
and for sources with shell, cometary, and irregular morphologies, we present integrated flux and peak intensity in an enclosed region above 
the 5$\sigma$ contours. Table~\ref{tab:7mm_sources} lists these observed values. In Appendix A, we show the derived properties of these regions, calculated using the assumptions and formulas in \cite{wood89}.

\begin{deluxetable*}{lllllll}[t!]
\tablecaption{Observed 7 mm (45.454.GHz) Continuum Parameters of Central Sources of W49A}
\tablewidth{700pt}
\tabletypesize{\scriptsize}
\tablehead{
\colhead{Source Name} & \colhead{RA} & \colhead{Dec} & \colhead{Peak Intensity}\tablenotemark{a}  & \colhead{S$_\nu$} & \colhead{Size, PA} & \colhead{Morphological}  \\ 
\colhead{} &  \colhead{(19h 10m)} & \colhead{(09$\degree$ 06$ \arcmin $)} &\colhead{(mJy beam$^{-1}$)} & \colhead{(Jy)} & \colhead{(\arcsec$\times$\arcsec, $\degree$)} &  \colhead{Type}
}
\startdata
A1 & 12.887 s & 12.12$\arcsec$   & 19.5 & 0.042$\pm$0.004 & $0.07\times$0.05, 99 & Spherical\\ \\
A2 & 12.890 s & 11.71$\arcsec$   & 18.8 & 0.252$\pm$0.003 & $0.66\times$0.50, 41 & Shell \\ 
B1  & 13.117 s & 12.33$\arcsec$   & 28.3 & 0.159$\pm$0.002 & $0.10\times$0.09, 148 & Spherical\\
B2 & 13.151 s & 12.67$\arcsec$   & 26.1 & 0.601$\pm$0.006 & $0.21\times$0.17, 48 & Shell\\
B3 & 13.154 s & 12.86$\arcsec$  & 16.4 & 0.131$\pm$0.001 & $0.17\times$0.07, 154 & Bipolar/Elongated\\ \\
C & 13.145 s & 18.74$\arcsec$  & 15.0 & 0.225$\pm$0.001 & $0.19\times$0.10 & Cometary\\ \\
D & 13.222 s & 11.162$\arcsec$  & 3.9 & 0.276$\pm$0.001 & $0.8\times$0.8 & Shell\\ \\
E1\tablenotemark{b} & 13.225 s & 12.14$\arcsec$  & 2.3 & 0.024$\pm$0.001 & $0.14\times$0.12, 14 & Spherical\\ \\
G1 & 13.377 s &  12.50$\arcsec$ & 8.2 & 0.108$\pm$0.004 & $0.37\times$0.45 & Shell\\
G2a & 13.419 s &  13.00$\arcsec$ & 31.3 & 0.376$\pm$0.004 & $0.15\times$0.13, 35 & Spherical\\
 G2b & 13.432 s  & 13.00 $\arcsec$ & 15.9 & 0.074$\pm$0.0021& $0.10\times$0.08, 30 & Spherical/Elongated\\
G2c & 13.456 s & 12.80$\arcsec$  & 28.5 & 0.148$\pm$0.002 & $0.14\times$0.06, 122 & Bipolar/Elongated\\
G3a & 13.494 s &  12.41$\arcsec$ & 3.1 & 0.025$\pm$0.004 & $0.32\times$0.19 & Cometary\\
 G3b & 13.494 s  & 11.76 $\arcsec$ & 3.1 & 0.030$\pm$0.0021& $0.24\times$0.14 & Cometary\\
G3c & 13.548 s & 11.70$\arcsec$  & 2.4 & 0.090$\pm$0.002 & $0.98\times$0.12 & Irregular\\
G3d & 13.596 s & 10.97$\arcsec$  & 2.0 & 0.057$\pm$0.002 & $0.75\times$0.14 & Irregular\\
G4 & 13.567 s & 13.75$\arcsec$  & 1.9 & 0.165$\pm$0.002 & 2.$12\times$0.27 & Irregular\\
\enddata
\tablenotetext{a}{The 5$\sigma$ error in these peak intensities is 0.8 mJy beam$^{-1}$}
\tablenotetext{b}{Sources E2 and E3 are not detected at the 3$\sigma$ (0.5 mJy beam$^{-1}$) level.}
\label{tab:7mm_sources}
\end{deluxetable*}

\subsubsection{Recombination Lines}
\begin{figure*}[t!]
\gridline{\fig{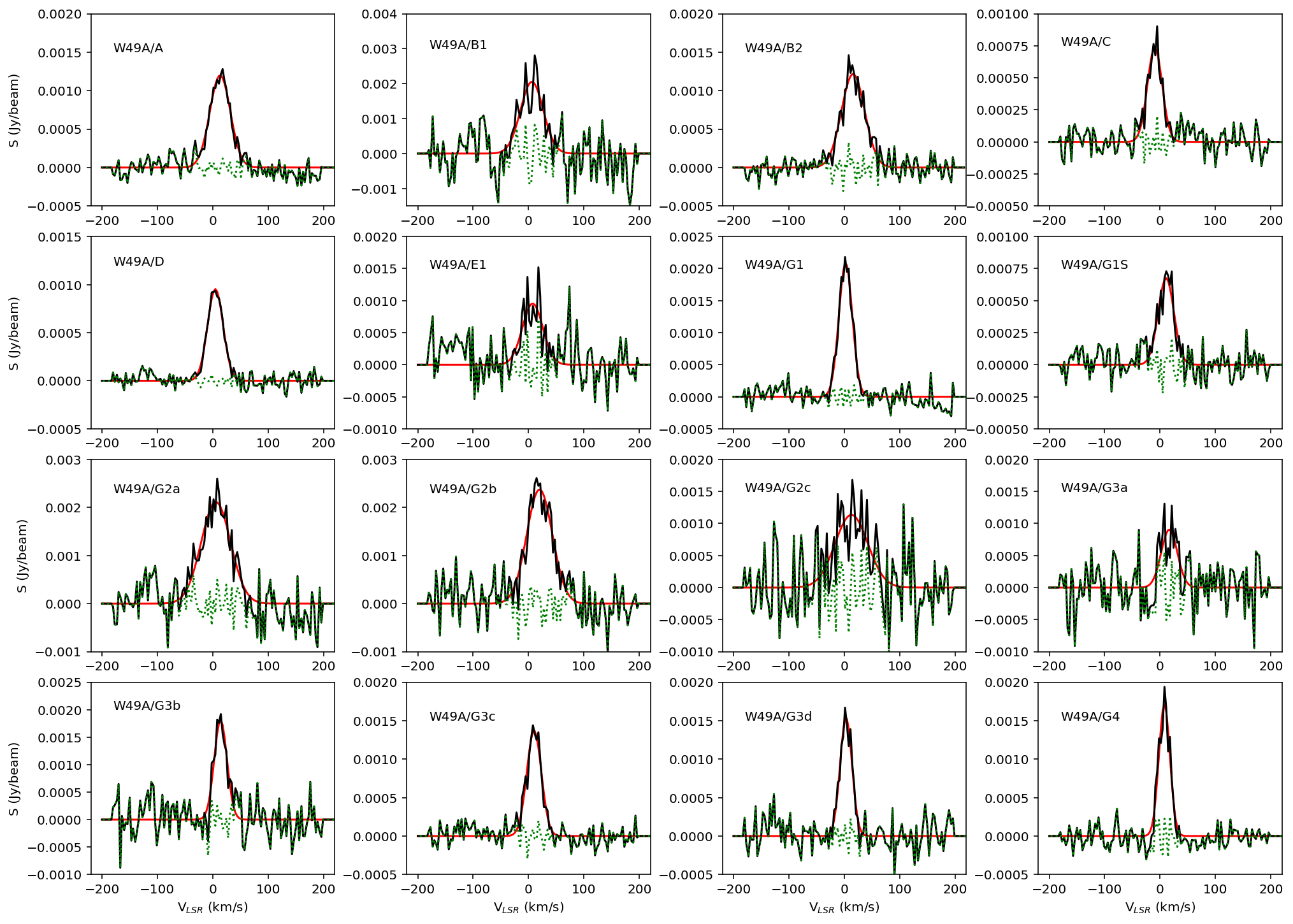}{0.9\textwidth}{}
}
\caption{Sources with detected H52$\alpha$ line emission in W49A. 
The data are shown as solid black lines, Gaussian fits as solid red lines, and residuals in dotted green lines. 
The rms noise in the H52$\alpha$ line data is 0.2 mJy beam$^{-1}$. 
}
\label{fig:7mmRRL} 
\end{figure*}

The 45.454 GHz sub-band centered on the H52$\alpha$ line was calibrated and imaged separately. These data were imaged and self-calibrated as described above, using the same solution intervals. The continuum was subtracted to produce a line-only image of the H52$\alpha$ emission.
The H52$\alpha$ observations have a velocity coverage of $\sim$420  km s$^{-1}$ (3.3 km s$^{-1}$ channel$^{-1}$). The $\it{rms}$ noise and other observing parameters of the 7 mm line data are given in Table~\ref{tab:VLAobs}.  Gaussian fits were made to each of the detected 7 mm sources using the SciPy Python library. The individual recombination line profiles, Gaussian fits and residuals of the sources with detected 7 mm line emission are given in Figure~\ref{fig:7mmRRL}. The fit parameters are given in Table~\ref{tab:RRL}.

\subsection{NOEMA 1.2~mm}

We observed W49A with NOEMA on 2015 August 20, 21, and 24 in the extended 7A6 configuration 
(maximum baseline of 760~m) and at a frequency of 257 GHz.  The spectral setup covered, in particular, the H29$\alpha$ line and
the CH$_3$CN $J$=14$_K$--13$_K$ ladder with the Wideband Express (WideX) correlator (4~GHz bandwidth per polarization, with 1024 $\times$ 3.9~MHz channels).
The unusual summer scheduling of this long baselines configuration was promoted to test sources amenable to
self-calibration, like W49A. The observations were performed under conditions of precipitable water vapor columns of 5 to 8~mm, 
with corresponding $T_{\rm sys}$ values of 500--1000 K. Individual scans were 10~s. Pointing was done every 30 to 60 min. 

These data were processed using the standard NOEMA pipeline as implemented in the GILDAS package CLIC.
Data reduction consisted of first auto flagging any deviant data points, then using the atmospheric phase monitor 
in each antenna to correct for antenna-based short-time phase variations. The bandpass was calibrated using observations 
of the complex gain calibrator, 1749+096, and then phase variations as a function of time were derived and applied. 
The flux was calibrated using either observations of  MWC349 (August 20 and 21) or 1749+096 (August 24). 
Lastly, the amplitude was calibrated, again using observations of 1749+096.
The bandpass, phase and amplitude solutions were inspected during calibration, and all looked satisfactory. 
The pipeline flux calibration worked well for the August 20 and 21 observations using MWC349. For August 24, the scatter was 
larger through the night, and again the stronger 1749+096 gain calibrator was used, taking the value for the flux density 
obtained from the earlier two nights. The flux density uncertainty is at least 20\%. Details of the observations are given in Table 1.

Because no proper baseline solution was available for the antenna configuration, special steps were taken in the phase calibration process. In particular, a baseline-based phase calibration was performed, rather than the default antenna-based calibration. Further precautions were taken in self-calibrating the continuum science data. Self-calibration was done over several iterations, going from a solution interval of 240 s and including only a few ($\sim$10) clean components, down to 60 s and cleaning the entire image. Only the phase was self-calibrated in each iteration. The rms noise was improved by a factor of almost two in this process, going from 42 mJy beam$^{-1}$ to 24 mJy beam$^{-1}$ in the final continuum image. The peak flux density in the image
was 0.63 Jy beam$^{-1}$ (source G), 
i.e. a maximum signal-to-noise ratio of $\sim$ 25, strongly limited by dynamic range considerations stemming  
from imperfect gain calibration and poorly sampled extended structure.

Three previously known sources (A, B2 and G2) were detected in the continuum, and fitted in CASA. The integrated flux densities of the continuum sources were A: 0.20$\pm$0.04 Jy, B2: 0.31$\pm$0.06 Jy and G2: 0.78$\pm$0.16 Jy. The peak intensities for the sources were A: 0.12$\pm$0.02 Jy beam$^{-1}$, B2: 0.31$\pm$0.02 Jy beam$^{-1}$ and G2: 0.63$\pm$0.02 Jy beam$^{-1}$. 
These values are in line with the higher resolution 1.4~mm imaging results of \citet{wilner2001} 
that indicate these sources are dominated by free-free emission with apparently low or 
            moderate
optical depth.

The phase solutions obtained from self-calibrating the continuum data were applied to the
            WideX
spectral data. 
The resulting rms noise is 50 mJy beam$^{-1}$ in 4.6 km s$^{-1}$ channels. 
The synthesized beam size for the images was 0\farcs6$\times$0\farcs2, with position angle (PA) of $-$160$^\circ$. 
We note that while the channel width is similar to the VLA observations (4.6 km s$^{-1}$), 
the H29$\alpha$ line was located toward the
            WideX
band edge, which limits the redshifted velocity coverage. We extracted the portion of
            the WideX spectrum bracketing the H29$\alpha$ line. The WideX
spectrum contains emission from many molecular lines, one of which is located close to the redshifted wing 
of the H29$\alpha$ line. 

\begin{deluxetable*}{lcccc}[t!]
\tablecaption{Radio Recombination Line Parameters at Common Angular Resolution\tablenotemark{a}}
\tablewidth{700pt}
\tabletypesize{\scriptsize}
\tablehead{
\colhead{Source Name} & \colhead{Recombination} & \colhead{Amplitude} & \colhead{V$_{LSR}$}  & \colhead{$\Delta V_{\rm Convl}$} \\ 
\colhead{} & \colhead{Line} & \colhead{(mJy beam$^{-1}$)} &\colhead{(km s$^{-1}$)} & \colhead{(km s$^{-1}$)}\\
(1) & (2) & (3) & (4) &(5)
}
\startdata
\hline\hline
A & H29$\alpha$ & 43$\pm$2 & 13.4$\pm$0.9 & 47.7$\pm$2.1 \\
& H52$\alpha$ & 68$\pm$8 & 15.6$\pm$2.1 & 37.2$\pm$5.0 \\
& H92$\alpha$ & 2.4$\pm$0.1 & 9.3$\pm$2.1& 78.9$\pm$5.1 \\
  \hline
B2& H29$\alpha$  & 83$\pm$3 & 11.1$\pm$0.8 & 52.1$\pm$1.8 \\
 & H52$\alpha$ & 33$\pm$6  & 16.5$\pm$5.5 & 63.9$\pm$13.0 \\
 & H92$\alpha$ & 2.9$\pm$0.2 & 7.2$\pm$1.8 & 51.9$\pm$4.3 \\
 \hline
G2 & H29$\alpha$ & 134$\pm$3 & 16.5$\pm$0.5 & 50.0$\pm$1.2 \\
 & H52$\alpha$ & 47$\pm$8  & 12.0$\pm$4.7 & 60.0$\pm$11.0 \\
 & H92$\alpha$ & 6.6$\pm$0.3 & 10.4$\pm$1.2 & 55.6$\pm$2.7 \\
\hline
\enddata
\tablenotetext{a}{These are the parameters of the three sources detected at all three frequencies, convolved to the lowest resolution, that of the VLA-BCD data. The line profiles for these sources are shown in Figure~\ref{fig:ThreeSourceRRL}}
\label{tab:ThreeSourceRRL}
\end{deluxetable*}

\begin{figure*}[t!]
\gridline{\fig{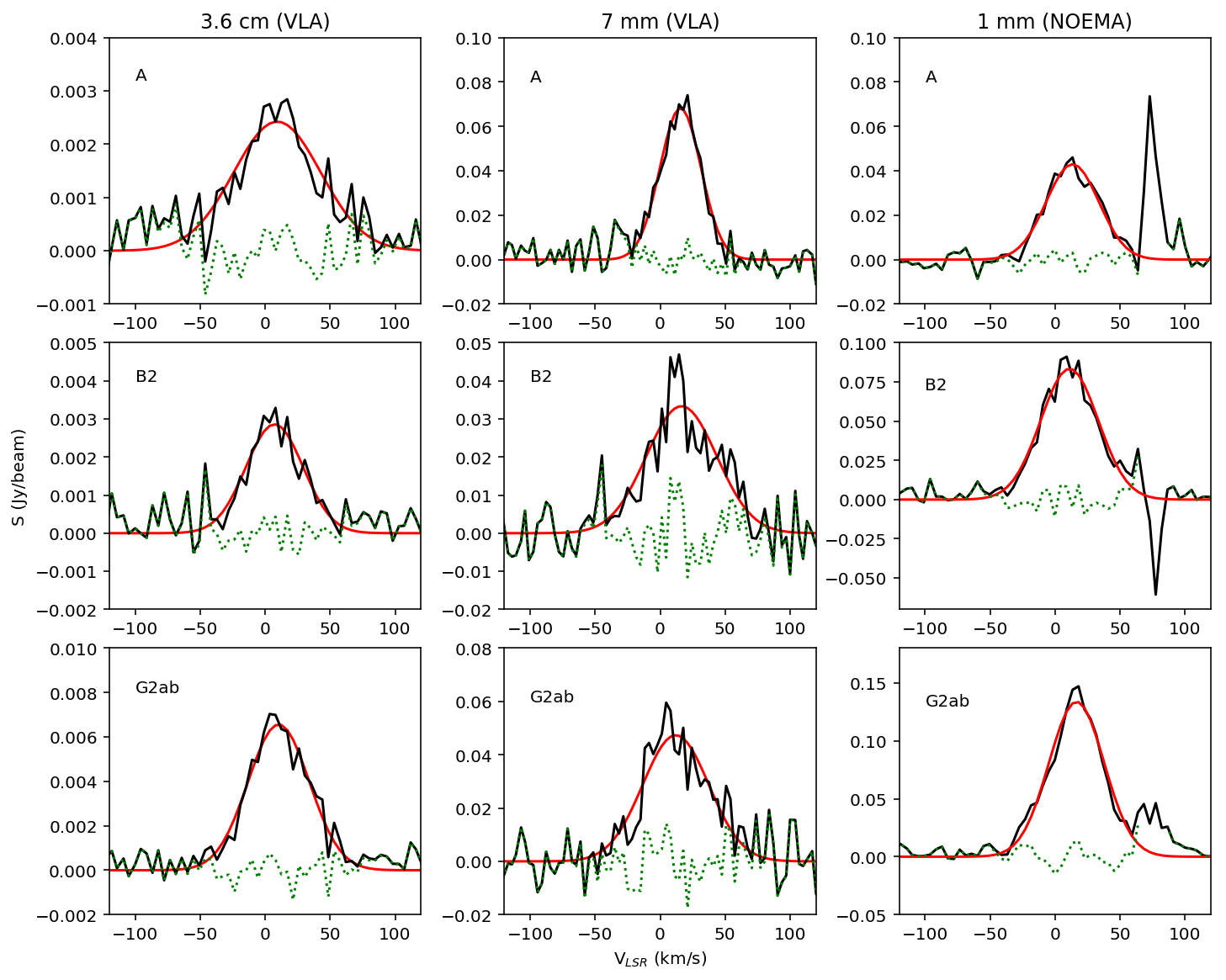}{0.8\textwidth}{}
        }
\caption{Recombination lines for sources A, B2 and G2ab at 3.6 cm (VLA), 7 mm (VLA) and 1 mm (NOEMA). These line profiles were made from data convolved to a common resolution and PA of 0\farcs95$\times$0\farcs75, $-$66\degree. Solid black line shows the data, solid red line is the Gaussian fit to the data, and the dashed green line shows the residual. Fits to the convolved 1 mm data for these three sources are shown in Table 4. The portion of the spectrum from $-$120 to 120 km s$^{-1}$ is shown for each line. We tentatively identify the feature on the redshifted wing of the H29$\alpha$ line as the 
molecular line CH$_3$C$_2$H (methyl acetylene), seen mainly in absorption at source B2, and in emission at sources A and G2.
}
\label{fig:ThreeSourceRRL} 
\end{figure*}

The 1.2 mm and 7 mm data were convolved to the resolution of the 3.6 cm VLA-BCD data, and line profiles were generated for each of the sources detected in common (A, B2 and G2). These profiles and the Gaussian fits to them are shown in Figure~\ref{fig:ThreeSourceRRL}. Table~\ref{tab:ThreeSourceRRL} provides the fits to the convolved recombination line data just for the three sources (A, B2 and G2) detected at 1.2 mm, 7 mm and 3.6 cm.

\section{RESULTS}

\subsection{Radio Continuum Source Morphologies}
These are the first observations of the W49A region at 3.6~cm with the VLA in the A-configuration. 
At this higher angular resolution ($\sim0\farcs15\approx0.008$~pc$~\approx1650$~au), 
the images reveal new source morphologies and new sources, described in detail below. 
As often noted in studies where low resolution observations are followed up at higher resolutions, many sources that appear 
similar at lower resolutions separate into a variety of distinct morphologies. 

\subsubsection{Source A}
At both low resolution (Figure~\ref{fig:3.6cmBCD}) and high resolution (Figure~\ref{fig:3.6cmA}), source A has an elongated, bipolar morphology. The new high resolution 3.6 cm image (Figure~\ref{fig:3.6cmA}) shows that the source has a clear edge-brightened double lobed structure, with the lobes extending $\sim$1\farcs9 to the NW and 0\farcs8 to the SE ($\sim$0.1 pc). At 7 mm, source A separates into A1 and A2, with A1 being the compact source to the NE. The continuum emission can be seen as contours in Figure~\ref{fig:7mmA}a. 
In A2, the 7~mm emission resembles a flattened ``disk'' with a central gap of diameter $\sim$0\farcs08 ($\sim$880~au), and a long axis perpendicular to the bipolar lobes seen at 3.6 cm. 

\subsubsection{Source B}
Source B has a complex substructure that can be seen in Figure~\ref{fig:3.6cmA} and Figure~\ref{fig:7mmA}b. The subsources in source B were detected previously \citep{dep2004}, but for the first time we see the B subsources at similar resolution at 7 mm and 3.6 cm. The B1 source, located to the SW, has an elongated (bipolar) morphology at 3.6 cm. This source is also elongated but more centrally peaked at 7 mm.
Source G2c (below) has a similar elongated morphology at 7 mm.

\subsubsection{Sources C and C1} The cometary sources C and C1 are seen in Figure~\ref{fig:3.6cmA} and Figure~\ref{fig:7mmA}c. Both have edge-brightened morphologies, with the bright faces pointed to the NW (C) and SW (C1). In both cases, the 7 mm emission traces the 3.6 cm emission, but with less sensitivity to faint structures. In sources that are optically thin and resolved, we expect the emission at 7 mm to be more faint. \cite{rod2020} have reported large proper motions for source C equivalent to a plane of the sky velocity of $\sim$76$\pm$6 km s$^{-1}$ in the direction of the bow shock face. 

\subsubsection{Source D} Source D is seen in Figure~\ref{fig:3.6cmA} and Figure~\ref{fig:7mmA}d to have a distinct edge-brightened shell-like structure with a diameter of $\sim$0\farcs8 (0.04 pc). At 7 mm, breaks in the shell on the E edge are apparent.

\subsubsection{Source E}
At both 3.6 cm and 7 mm, the previously known source E, located to the NNE of source D, is resolved into 3 sub-sources (E1, E2 and E3) as shown in Figure~\ref{fig:7mmA}d.
At the resolution of the 3.6 cm observations, the 3 sources appear spherical. At the higher resolution of the 7 mm observations, 
it appears that the E1 source remains centrally brightened, while E2 and E3 may be edge-brightened shells with diameters of $\sim$0\farcs2 (0.01 pc), appearing to be smaller versions of source D.

\subsubsection{Source F} Source F is seen in Figure~\ref{fig:3.6cmA} and Figure~\ref{fig:7mmA}e at 3.6 cm to have a cometary morphology with the bright face to the W/SW. At 7 mm, there is only a single contour visible at the 5$\sigma$ level.
 
\subsubsection{Source G}
The source G region is crowded and resolves into complex substructures. 
Some of the complex source morphologies previously imaged at 7 mm were discussed in \cite{dep2000}. The new high resolution 3.6 cm image shows many of 
these same morphologies.


Spherical at 3.6 cm, G1 and G1S are observed at 7 mm (more optically thin) to be edge-brightened shells, with source G1 having a more elongated shape. The G2 subsources can be seen in Figure~\ref{fig:7mmA}f, and are labeled G2a, G2b and G2c. G2c in particular has a distinct, elongated structure, extending from SE to NW, with stronger emission at 7 mm than 3.6 cm.
Source G3 appears to also break up into subsources in the high resolution 3.6 cm data, seen in Figure~\ref{fig:7mmA}g. 
Sources G3a and G3b are cometary, with their cometary arcs facing the G2 sources. source G4 and G5 are more complex, composed of several interconnecting arcs and filaments.
We note that in Figure~\ref{fig:3.6cmA}, the high resolution 3.6 cm image, G3, G4 and G5 appear make up the edges of a more coherent cavity structure of size $\sim$ 3\farcs5, extending to the east from source G1 and G2. These three sources do not contain the bright, compact structures found in G1 and G2.

\subsection{Recombination Line Results}
We have observed radio recombination line emission at 3.6 cm (H92$\alpha$), 7 mm (H52$\alpha$) and 1.2 mm (H29$\alpha$). 
The line profiles at 3.6 cm and 7 mm are shown in Figures~\ref{fig:H92alpha} and \ref{fig:7mmRRL}. 
Gaussian fits were made to each detected line, and the parameters of these fits (amplitude, LSR line velocity, and linewidth) for each detected source are given in Table~\ref{tab:RRL}. 
In sources where the line was detected in both previous observations \citep{dep1997} and the new observations,
we list the previously published linewidth and the ratio of the fitted linewidths. 
For the three sources with line detections at three wavelengths (A, B2, G2), we convolved the H29$\alpha$ and H52$\alpha$ line data to the resolution of 
the  H92$\alpha$ data ($\sim$0\farcs8). Figure~\ref{fig:ThreeSourceRRL}  shows these spectra, Gaussian fits, and residuals.  The parameters from these 
fits are listed in Table~\ref{tab:ThreeSourceRRL}.

We tentatively identify the feature on the redshifted wing of the H29$\alpha$ line as the molecular line CH$_3$C$_2$H (methyl acetylene) 
after examining line lists for this frequency range in the “Splatalogue” compilation \citep{remijan2007}, as well as identifications 
in other, similar, star-forming regions \citep[e.g.][]{klaassen2018}. This line is seen mainly in absorption at source B2, and in emission at sources A and G2. 
The H29$\alpha$ line is likely blended with emission from additional molecular lines, although the similarities in line shape with H52$\alpha$ 
suggest the impact cannot be very important for these sources. \citet{klaassen2018} describe a procedure to separate the molecular and recombination lines that incorporates information from the surrounding molecular emission. Unfortunately, the severe spatial filtering and modest dynamic range 
of the NOEMA data preclude the application of this procedure here.

\subsection{Linewidths and Velocities}
We have used the line parameters at 3.6~cm and 7~mm to determine the kinematic linewidth 
and electron density for each source detected at both wavelengths, following the method outlined in \cite{keto2008}. From their
Equation (1), thermal and kinematic linewidth are assumed to add in quadrature to produce the observed high frequency (in this case H52$\alpha$) Gaussian line. Assuming an electron temperature of 10,000 K, and the observed H52$\alpha$ linewidth, we determine the kinematic linewidths. These widths ($\Delta V_{\rm Kin}$) are shown in Table~\ref{tab:RRL}. Electron densities were calculated according to
Equation (2) in \cite{keto2008}, using the detected widths of the low (H92$\alpha$) and high (H52$\alpha$) frequency lines as $\Delta \nu_V$ and $\Delta \nu_G$ respectively to solve for $\Delta \nu_L$. This value of $\Delta \nu_L$ was then used in their Equation (3) to derive the electron density.
The derived electron densities vary from 0.2 to 1.7$\times$ 10$^5$ cm$^{-3}$, and are also reported in Table~\ref{tab:RRL}. \cite{r-s2020} also found electron densities of $\sim$10$^4$ to 10$^5$ cm$^{-3}$ for the majority of HII regions with diameter $<$0.05 pc in W51A, which is in the same cluster regime as W49A.  In these two regions, high electron density ($n_e > $10$^6$ cm$^{-3}$) HC~\ion{H}{2} regions appear to be rare.

In general, the sources with the most compact subcomponents tend to show broader kinematic linewidths, with G2a, G2b and G2c all having linewidths $>$50~km~ s$^{-1}$.
For sources A, B2, and G2,  the center velocity of the 3.6~cm line is blueshifted by  $\sim5$--10~km~s$^{-1}$ from the center velocities 
of the 7~mm and 1.2~mm lines when viewed at common angular resolution (Table~\ref{tab:ThreeSourceRRL}). 
Similarly, in W51A, \cite{r-s2020} found that the H77$\alpha$ line at 2 cm is blueshifted by a few km s$^{-1}$ with respect to the H30$\alpha$ line at 1.3 mm for the seven small HII regions with detections in both lines. 
Such shifts are expected in a scenario where the ionized gas is both expanding and partially optically thick, 
as discussed in detail by \cite{peters2012}. These three sources have spectral indices that indicate that they are partially optically thick between 8.3 and 43 GHz (A and B2), and between 22 and 43 GHz (G2) \citep{yang2018}.
The unusual rising spectral indices to shorter wavelengths in these sources \citep{dep2004} are also a natural outcome of this configuration \citep{keto2008}.

\section{Discussion}
\subsection{3.6 cm Recombination Line Properties}
The H92$\alpha$ line at 3.6~cm was observed in the B, C, and D configurations of the VLA in both 1994--1995 and 2015--2016 with effectively the same angular resolution.
This allows for a direct comparison over a 21 year time baseline. A caveat is that the newer data have higher sensitivity and broader 
velocity coverage, so line parameters obtained from single Gaussian fits may not be in agreement, even with no changes in the source emission. 

In general, we find that single Gaussian fits to the new data tend to have slightly broader linewidths, as recombination line profiles from \ion{H}{2} regions are not 
perfectly Gaussian, and the line wings tend to be better detected in the new observations.  As reviewed by \cite{peters2012}, the line profile in an \ion{H}{2} region is a convolution of a Gaussian profile (from thermal and microturbulent broadening) and a Lorentzian profile (caused by electron pressure broadening), resulting in a Voight profile. Thus, these line wings become 
more prominent in sources where pressure broadening is a significant effect.

In some sources, the single Gaussian fits appear to leave a residual at the line peak. This effect can be seen in Figure~\ref{fig:H92alpha}, e.g. for source C and source D, where the narrower line peak that was fit more 
closely in the 1994--1995 data is clearly visible in the residual of the Gaussian fit to the 2015--2016 data.
   Such narrow peaks were predicted in a semi-analytic model of a bipolar outflow
   observed from close (30$\degree$) to its axis by \citet{tan2016}. However, these narrow residuals are only at the 2$\sigma$ level.
   
In source D, as a result of the residual at the peak, the fitted 
FWHM is broader, 58.7$\pm$2.1 km s$^{-1}$ in 2015--2016, compared to 48.1$\pm$1.6 km s$^{-1}$ reported for the 1994--1995 data.
For sources where this narrower component is not apparent, e.g source B (source B2 in the current work), the fitted FWHM parameters are very similar, 
55.4$\pm$6.1 km s$^{-1}$ compared to 56.4$\pm$5.4 km s$^{-1}$ reported for the 1994--1995 data.
A few sources appear to have narrower fitted linewidths in the 2015--2016 data, e.g. for source E, the fitted FWHM is  57.3$\pm$3.1 km s$^{-1}$ 
compared to 64.8$\pm$4.7 km s$^{-1}$ reported for the 1994--1995 data, a 2$-$3 $\sigma$ difference of~$-$7.5$\pm$5.6 km s$^{-1}$. In addition, marginal detections in the 1994--1995 data were likely more uncertain than suggested by the reported formal errors. 
For example, source C, 
with signal-to-noise ratio of $\sim$4, had a reported linewidth of 18.6$\pm$2.8 km s$^{-1}$, apparently a significant underestimate of the true linewidth of this source, measured to be 57.0$\pm$3.3  km s$^{-1}$ in the current observations. 

In the case of sources C and D, there is the suggestion of asymmetrically blueshifted peak in the line profile (see Figure~\ref{fig:H92alpha}). 
This effect is predicted by \citet[see their Figure 9]{peters2012} to be the result of a redshifted absorption shoulder from an optically thick region at lower frequency.\footnote{We note, however that their prediction is based on spatially resolved lines, and these line profiles have been averaged over each region.} The source flux densities of sources C and D at 22.2 GHz \citep{dep2000} and 8.3 GHz \citep{dep1997} can be used to determine a spectral index ($S_\nu\propto\nu^\alpha$) of 0.4 and 0.1 respectively, indicating that these two sources are marginally optically thick at 3.6 cm.
The asymmetry will occur in an expanding, optically thick UC~\ion{H}{2} region at sufficiently high frequency if pressure broadening does not overwhelm the effect  (see \citealt{peters2012} Figure~14).  For expansion at the sound speed of ionized gas, they predicted that this would only occur at wavelengths shortward of 1~cm.

Detection of an effect at 3.6~cm thus would require superthermal expansion velocities.  This appears consistent with the kinematic linewidths of 25--27~km~s$^{-1}$ found for these sources---well more than twice the thermal value. However, the asymmetry would be expected to be more pronounced at shorter wavelengths, which does not appear consistent with the more symmetric H52$\alpha$ line shapes shown in Figure~\ref{fig:7mmRRL}. We do note, though, that looking for these subtle changes in line shapes between these data at different frequencies is complicated by the different native resolutions of the three data sets. The H92$\alpha$ data sets being compared between the two epochs are matched in observing configuration and thus very close in native resolution.

\subsection{Newly Detected Broad lines in Source A}
While radio recombination line emission was not previously detected towards source A \citep{dep1997},
the new data show clear (and broad) line emission from this source at 3.6~cm, 7~mm, and 1.2~mm (see Figure~\ref{fig:ThreeSourceRRL}). 
There appears to be significant additional broadening in the line at 3.6 cm,  but the large derived kinematic width (39.2 km s$^{-1}$) indicates high velocity ionized gas motions of some sort are present, perhaps outflow along the long NW to SE axis.
An ionized outflow in source A would be consistent with the oppositely directed, edge-brightened lobes visible in Figures~\ref{fig:3.6cmBCD} and~\ref{fig:3.6cmA}.

\subsection{Linewidth Changes over 21 Years}
\subsubsection{H92$\alpha$ Comparison}
\begin{figure*}[t!]
\gridline{\fig{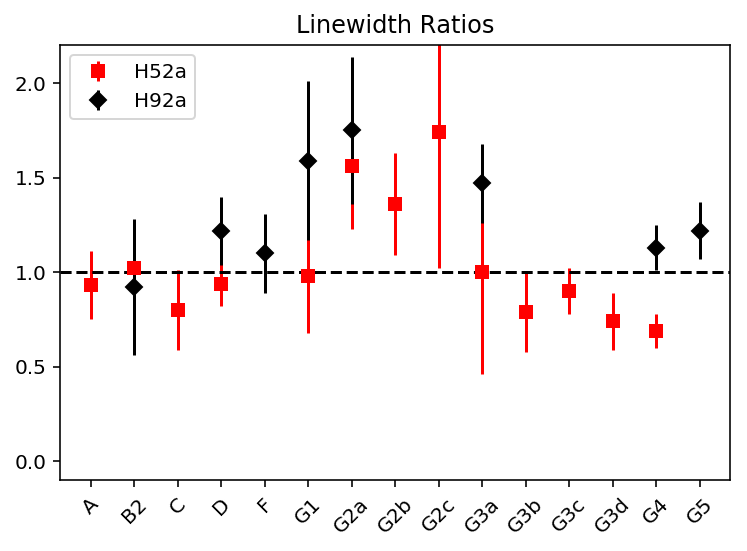}{0.6\textwidth}{}
}
\caption{The linewidth ratios from the two epochs in Table 2 for all detected sources. A ratio of 1 is indicated by the horizontal line. The 3.6 cm ratios (black diamonds) and 7 mm ratios (red squares) are shown with 3 $\sigma$ error bars. This plot indicates that most sources have a ratio of $\sim$1, indicating no significant change over the 21 year time baseline. One source (source C, not plotted) has a high ratio at H92$\alpha$ (3.6~cm) from a poor fit to the early epoch data, discussed in the text. The remaining sources that have ratios of $\sim$1.5 are G1 and G3 (only in the H92$\alpha$ line) and G2. G2 (and sub-sources) is the only region with  an increased linewidth ratio at both frequencies.
}

\label{fig:WidthRatio} 
\end{figure*}

The ratio of the measured linewidths at the two epochs discussed above and listed in Table~\ref{tab:RRL} are plotted in Figure~\ref{fig:WidthRatio}, with error bars indicating 3x the formal errors in the velocity ratio.
In Figure~\ref{fig:WidthRatio}, a ratio of 1 is indicated by the horizontal dotted line. The H92$\alpha$ ratios (black diamonds) and H52$\alpha$ ratios (red squares) indicate that most sources have a ratio of $\sim$1, with no significant change over the 21 year time baseline. The remaining sources that have ratios of $\sim$1.5 are G1 and G3 (only in the H92$\alpha$ line) and G2. G2 (and its sub-sources) is the only region with  an increased linewidth ratio at both detected frequencies over the 21 year time baseline.
Note that this plot excludes the high ratio for the H92$\alpha$ line from source C, as the apparent change in linewidth for this
source is likely due to incorrectly fitting a narrow peak in the low significance H92$\alpha$ detection at the 1994--1995 epoch. 
As discussed below, the H52$\alpha$ line in this source does not show a similar change in linewidth over this timescale. 

Substantial changes in linewidth over the 21 year time baseline at 3.6 cm are found for sources G1, G2 and G3. 
Source G1 increased in linewidth from 29.3$\pm$2.0 to 46.5$\pm$2.4 km s$^{-1}$ or +17.2$\pm$3.1 km s$^{-1}$. In source G2, the linewidth increased from 
31.7$\pm$1.8 km s$^{-1}$ to 55.6$\pm$2.7 km s$^{-1}$, or $+$23.9$\pm$3.2 km s$^{-1}$. These large changes do not appear to be influenced by the 
shape of the line, or by low signal-to-noise data, and a single Gaussian appears to provide a good fit at both epochs. For sources G1 and G3, we do not detect a similar change in the H52$\alpha$ line (see below).

We note that the change in linewidth for source G2 has occurred over the same period that the continuum flux density for this source decreased by 20\% in peak intensity 
and 40\% in integrated flux \citep{dep2018}. Both the continuum and the line data indicate a substantial and contemporaneous change in the ionized gas over this time period as discussed in more detail below in Sect.~\ref{sub:evolution-G}.
  Although in principle a substantial decrease in temperature over this period could have produced such effects \citep[Eq.~10]{brown1978}, in practice no mechanism exists that could explain such a large, sudden temperature drop.
    
\subsubsection{H52$\alpha$ Comparison}
The observations of the higher frequency H52$\alpha$ line support the results obtained from the H92$\alpha$ line, even as the mismatched 
angular resolution at 7~mm from 1995 and 2016 makes the comparison less direct ($\sim$1\farcs65 vs 0\farcs04). 
The ratio of the measured linewidths at the two epochs discussed above and listed in Table~\ref{tab:RRL} are plotted in Figure~\ref{fig:WidthRatio}. 
For sources A, B, C and D, the H52$\alpha$ linewidths show no significant changes from the values reported by \citet{dep1997}. 
For example, the linewidth of source A is effectively constant, 43.6$\pm$1.7 km s$^{-1}$ in 2016 compared to 46.8$\pm$2.2 km s$^{-1}$ in 1995. 
For source G2, the H52$\alpha$ linewidths in the various subsources range from 55$\pm$3 to 70$\pm$9 km s$^{-1}$, with an average value of 
60$\pm$10 km s$^{-1}$. This can be compared to the previous observation G$_{West}$, which would have been dominated by the subsources G2a, G2b, and G2c,
with a reported linewidth of 40.4$\pm$1.1 km s$^{-1}$. So at 7~mm as well, the G2 source has undergone a significant increase in its linewidth of $+$20$\pm$10 km s$^{-1}$ from 1995 to 2016. 
Therefore, the linewidth of source G2---and no other sources detected at 7 mm at both epochs---increased between the observations 
made  in 1995 and 2016. 

We note that the H52$\alpha$ line was also observed with the VLA in the A configuration in 2001, albeit at much lower spectral resolution
and with less velocity coverage, and is described in \cite{dep2004}. These 2001 observations span a shorter time baseline than provided by both the 3.6 cm (1993-4) and 
lower angular resolution 7 mm (1995) data. In addition, the recombination lines of the G2 subsources observed in 2001 all had low signal to noise ratios.
Even in these lower quality spectra, the high angular resolution 2001 observations provided a hint that the linewidths of the G2 subsources might have broadened between 
1995 and 2001, unlike those of the other sources in the region (see Table~1 of \citeauthor{dep2004} \citeyear{dep2004}). 

\subsection{Evolution of Source G}
\label{sub:evolution-G}
As noted in \citet{smith2009} and \citet{dep2018}, the source G region contains many of the most compact, highest emission measure sources, 
which are presumably very young.
   Indeed, \citet{gmr1992} found that water masers trace an expanding bipolar outflow in this region that \citet{mac1994} showed was consistent with a jet coccoon having an age of roughly 350~yr at that epoch.
Since these sources are expected to be actively accreting and ejecting material in the scenario of \citet{peters2010a},
they are the most likely to experience changes in flux density. 
\cite{dep2000} reported that sources G2a and G2b have rising spectral indices between 13~mm and 3.3~mm, another feature  predicted in the hydrodynamic simulations of gravitationally unstable accretion \citep{peters2010a}. 
Rising spectral indices are typically associated with ionized outflow.

\cite{dep2018} noted that the position of the detected flux density decrease at 3.6 cm to the east of G2 is slightly offset from the continuum peak and 
aligns with the G2b cavity proposed in \citet{smith2009}. 
It is possible that a decrease in the brightness of the G2b cavity results from the protostar moving into a region of higher density and ionizing a smaller volume of gas. Furthermore, the elongated shape of G2c (seen in the new high-resolution data at 7 mm and 3.6 cm) could indicate the presence of a bipolar ionized outflow from a high mass protostar. The ionization of a smaller, higher density region of gas closer to the star would be in agreement with the detected increase in linewidth (at 7 mm and 3.6 cm)  from this source over the same timescale that the flux density has decreased.
   \citet{mcgrath2004} carefully calibrated the locations of the water maser outflow
   and the continuum emission associated with the G2 source.
%

The central position of the water maser outflow measured by \citet{gmr1992} and \citet{mcgrath2004} is RA 19h10m13.415s, Dec 9$\degree$ 6$\arcmin$ 12\farcs92 (J2000), close to the peak of source G2a. This position in indicated by a black star in Figure~\ref{fig:7mmA}f. 
This outflow extends over an ellipse with semi-major axis of $\sim$1\arcsec, covering the
   entire
G2 region and even a bit beyond.
            The outflow shows a three-dimensional velocity distribution consistent with a Hubble flow expanding from a point: velocity linearly increasing with radius.
This kinematic behavior can be explained by the masers forming in the shocked thin shell swept up by the
   expanding
cocoon of a
protostellar jet \citep{mac1994},
            in which material on the sides, closer to the source, moves more slowly.
Outflows driven by the inner disks were not included in the models of \citet{peters2010a}, although they were later studied in the same simulation framework by \citet{peters2014}.
   \citet{tan2016} did model radio recombination line formation in a bipolar outflow, but they did not
   follow secular evolution of this dynamic system, which appears likely to explain
   the observed linewidth changes.  
   
   We also note that sources G2 and A bracket the global column density peak of the entire W49N molecular clump with $>$10$^{5}$~M$_{\odot}$ of gas located within just a few pc \citep{gm2013,dep2018}.

\subsection{Source Monitoring}
The short timescale variability detected in both continuum flux density and recombination line parameters in source G2 indicates 
the need to monitor this source and others like it on a regular basis so that the changes can be characterized and 
utilized in conjunction with numerical models to improve our understanding of the early evolution of the youngest massive protostars. 
In addition, ALMA observations at shorter wavelengths with similar high resolution will penetrate deeper into the smallest and most optically thick \ion{H}{2} regions, providing new morphological and kinematic clues to the 
evolution of these smallest UC \ion{H}{2} regions.

\section{CONCLUSIONS}
We have obtained new VLA high resolution continuum and radio recombination line observations of W49A at 7 mm and 3.6 cm, supplemented by observations with
NOEMA at 1.2~mm. The 3.6~cm A configuration image, with $\theta\sim$0\farcs15, reveals new source morphologies, notably an edge-brightened 
bipolar structure eminating from source A, and 3 subsources within the source E. The three sources detected at the shortest wavelength,
(A, B2, and G2) all have kinematic linewidths in excess of $\sim$40~km~s$^{-1}$, indicating supersonic motions perhaps associated with youth.
Source G2, which is the only source in the region to undergo a significant flux density change between 1995 and 2015 (20\% decrease in peak intensity 
and 40\% decrease in integrated flux), also shows 
a significant increase in linewidth at both 3.6 cm and 7~mm over a similar 20 year time span.
This may be related to the evolution of the young bipolar outflow traced by the water masers in the region.

We plan to continue to make regular observations of the central W49A region at high resolution to search for further variations in flux density and gas kinematics.

\section*{Acknowledgements}
The authors thank the anonymous referee for providing helpful comments that improved this manuscript. CGD acknowledges support from NSF-RUI grant AST16-15311. The research of LEK is supported by a research grant (19127) from VILLUM FONDEN.
RSK acknowledges financial support from the German Research Foundation (DFG) via the Collaborative Research Center (SFB 881, Project-ID 138713538) 'The Milky Way System' (subprojects A1, B1, B2, and B8). He also thanks for funding from the Heidelberg Cluster of Excellence STRUCTURES in the framework of Germany's Excellence Strategy (grant EXC-2181/1 - 390900948) and for funding from the European Research Council via the ERC Synergy Grant ECOGAL (grant 855130). RGM acknowledges support from UNAM-PAPIIT project IN104319.
    M-MML was partly supported by NSF grant AST18-15461.
    
\appendix
\section{Derived 7 mm Continuum Properties}
Previous investigations of W49A and other massive star forming regions have tabulated the derived properties of their sub-regions, assuming a homogeneous, uniform density model. For completeness, we provide this same information here. 

Table~\ref{tab:7mm_properties} lists derived properties of the sources, using the observed 7 mm (45.454 GHz) continuum parameters from Table~\ref{tab:7mm_sources}, and the equations given in \cite{wood89}. Derived properties include radius, electron density ($n_e$), emission measure (U), mass of ionized gas ($M_{HII}$), log number of Lyman continuum photons ($log[N_{LyC}]$) , and equivalent zero age main sequence (ZAMS) spectral type for the ionizing source. We note that Tables~\ref{tab:7mm_sources} and~\ref{tab:7mm_properties} contain only sources detected at 7 mm, and thus some of the 3.6 cm detections are not listed here.

\begin{deluxetable*}{lccccccc}[t!]
\tablecaption{Derived Continuum Parameters of Central Sources of W49A from 7 mm Continuum}
\tablewidth{700pt}
\tabletypesize{\scriptsize}
\tablehead{
\colhead{Source Name} & \colhead{Size} & \colhead{n$_e$} & \colhead{EM} & \colhead{U}  & \colhead{M$_{H II}$} & \colhead{log[N$_{LyC}$]}\tablenotemark{a}  & \colhead{ZAMS}\tablenotemark{a,b}  \\ 
\colhead{} & (pc) &  \colhead{(10$^6$ cm$^{-3}$)} & \colhead{(10$^9$ pc cm$^{-6}$)} &\colhead{(pc cm$^{-2}$)} & \colhead{(10$^{-2}$ M$_{\odot}$)} & \colhead{(s$^{-1}$)} &  \colhead{Type}\\
(1) & (2) & (3) & (4) &(5) & (6) & (7) & (8)}
\startdata
A1 & 0.002 & 1.2 & 7.0 & 27.3 & 0.16 & 47.8 &  B1 \\
A2 & 0.02 & 0.1 & 0.44 & 49.7 & 11.9 & 48.6 &  O9 \\ 
\\
B1 & 0.004 & 1.1 & 10.0 & 42.6 & 0.64  & 48.4 & O9.5  \\
B2 & 0.008 & 0.8 & 9.8 & 66.4 & 3.5  & 48.9 & O8  \\
B3 & 0.004 & 0.8 & 6.4 & 40.0 & 7.1  & 48.3 & B0  \\
\\
C & 0.006 & 0.8 & 6.9 & 47.9 & 1.3  & 48.5 & O9  \\
\\
D & 0.03 & 0.06 & 0.3 & 51.2 & 20.0  & 48.6 & O9  \\
\\
E1 & 0.005 & 0.3 & 0.8 & 22.7 & 0.04  & 47.6 & B1.5  \\
\\
G1 & 0.02 & 0.1 & 0.4 & 37.5 & 4.6  & 48.2 & B0  \\
G2a & 0.006 &1.0 & 11 & 56.8 & 1.7  & 48.8 & O8.5  \\
G2b & 0.004 & 0.85 & 5.4 & 33.0 & 0.40  & 48.1 & B0  \\
G2c& 0.004 & 1.2 & 10.0 & 41.6 & 0.58  & 48.4 & O9.5  \\
G3a & 0.01 &0.1 & 0.2 & 23.0 & 1.1  & 47.6 & B1.5  \\
G3b & 0.01 & 0.15 & 0.38 & 24.4 & 0.9  & 47.7 & B1  \\
G3c& 0.01 & 0.07 & 0.15 & 24.4 & 1.89  & 47.7 & B1  \\
G3d & 0.01 & 0.1 & 0.3 & 30.3 & 2.4  & 47.9 & B0.5  \\
G4& 0.03 & 0.05 & 0.17 & 43.2 & 14.6 & 48.4 & O9.5  \\ \hline\hline
\enddata
\tablenotetext{a}{We note that these values are necessarily lower limits, since many of these HC/UC HII regions are partially optically thick, even at 7 mm.}
\tablenotetext{b}{As specified in Vacca et al. (1996)}
\label{tab:7mm_properties}
\end{deluxetable*}

\newpage
\software{CASA (McMullin et al. 2007), GILDAS (Pety 2005, GILDAS team 2013)}

\end{document}